\pdfoutput=1

\documentclass[twocolumn]{aastex62}

\graphicspath{{./}{figures/}}
\usepackage{import}
\usepackage{CJK}
\usepackage{xcolor}
\usepackage{savesym}
\savesymbol{tablenum}
\usepackage{siunitx}
\restoresymbol{SIX}{tablenum}
\usepackage{amsmath}

\submitjournal{ApJ}
\accepted{February 4th, 2022}

\begin{document}
\begin{CJK*}{UTF8}{gbsn}

\title{Physical Properties of the Host Galaxies of Ca-rich Transients}

\newcommand{\Purdue}{\affiliation{Department of Physics and Astronomy, Purdue University, 525 Northwestern Avenue, West Lafayette, IN 47907, USA}}

\newcommand{\IDSI}{\affiliation{Integrative Data Science Initiative, Purdue University, West Lafayette, IN 47907, USA}}

\newcommand{\PSU}{\affiliation{Department of Astronomy \& Astrophysics, The Pennsylvania State University, University Park, PA 16802, USA}}

\newcommand{\ICDS}{\affiliation{Institute for Computational \& Data Sciences, The Pennsylvania State University, University Park, PA, USA}}

\newcommand{\IGC}{\affiliation{Institute for Gravitation and the Cosmos, The Pennsylvania State University, University Park, PA 16802, USA}}

\newcommand{\MSU}{\affiliation{Department of Astronomy and Astrophysics, Michigan State University, 567 Wilson Road, East Lansing, MI 48824, USA}}

\newcommand{\NU}{\affiliation{Center for Interdisciplinary Exploration and Research in Astrophysics (CIERA) and Department of Physics and Astronomy, Northwestern University, Evanston, IL 60208, USA}}

\newcommand{\Berkeley}{\affiliation{ Department of Astronomy, University of California, Berkeley, CA 94720-3411, USA}}

\newcommand{\Carnegie}{\affiliation{The Observatories of the Carnegie Institution for Science, 813 Santa Barbara St., Pasadena, CA 91101, USA}}

\newcommand{\CalTech}{\affiliation{TAPIR, Walter Burke Institute for Theoretical Physics, 350-17, Caltech, Pasadena, CA 91125, USA}}

\newcommand{\OSU}{\affiliation{Center for Cosmology and AstroParticle Physics (CCAPP), The Ohio State University, 191 W. Woodruff Avenue, Columbus, OH 43210, USA.}}
\author[0000-0002-9363-8606]{Yuxin Dong (董雨欣)}
\Purdue
\NU

\author[0000-0002-0763-3885]{Dan Milisavljevic}
\Purdue
\IDSI

\author[0000-0001-6755-1315]{Joel Leja}
\PSU
\ICDS
\IGC

\author[0000-0002-4781-7291]{Sumit K. Sarbadhicary}
\MSU
\OSU

\author[0000-0002-2028-9329]{Anya E. Nugent}
\NU

\author[0000-0003-4768-7586]{Raffaella Margutti}
\Berkeley

\author[0000-0003-1103-3409]{Wynn V. Jacobson-Gal\'{a}n}
\Berkeley

\author[0000-0002-1633-6495]{Abigail Polin}
\Carnegie
\CalTech

\author[0000-0003-0776-8859]{John Banovetz}
\Purdue

\author{Jack M. Reynolds}
\Purdue

\author[0000-0001-8073-8731]{Bhagya Subrayan}
\Purdue

\begin{abstract}

Calcium-rich (Ca-rich) transients are a new class of supernovae (SNe) that are known for their comparatively rapid evolution, modest peak luminosities, and strong nebular calcium emission lines. Currently, the progenitor systems of Ca-rich transients remain unknown. Although they exhibit spectroscopic properties not unlike core-collapse Type Ib/c SNe, nearly half are found in the outskirts of their host galaxies that are predominantly elliptical, suggesting a closer connection to the older stellar populations of SNe Ia. In this paper, we present a compilation of publicly available multiwavelength observations of all known and/or suspected host galaxies of Ca-rich transients ranging from FUV to IR, and use these data to characterize their stellar populations with \textsc{prospector}. We estimate several galaxy parameters including integrated star formation rate, stellar mass, metallicity, and age. For nine host galaxies, the observations are sensitive enough to obtain nonparametric star formation histories, from which we recover SN rates and estimate probabilities that the Ca-rich transients in each of these host galaxies originated from a core-collapse vs.\ Type Ia-like explosion. Our work supports the notion that the population of Ca-rich transients do not come exclusively from core-collapse explosions, and must either be only from white dwarf stars or a mixed population of white dwarf stars with other channels, potentially including massive star explosions. Additional photometry and explosion site spectroscopy of larger samples of Ca-rich host galaxies will improve these estimates and better constrain the ratio of white dwarf vs.\ massive star progenitors of Ca-rich transients.

\end{abstract}


\section{Introduction}\label{sec:intro}

Understanding the progenitor systems of Calcium-rich (Ca-rich) transients is an active research topic in time-domain astrophysics. Compared to Type Ia and core-collapse (Type II, IIb, Ib/c) supernovae (SNe), Ca-rich transients evolve more rapidly into the nebular phase of emission \citep{2012ApJ...755..161K}. At early epochs, their spectra strongly resemble those of stripped-envelope Type Ib/c SNe, showing prominent lines of He I, O I, Mg II, and Ca II with large photospheric velocities ($\approx$ 11,000 km s$^{-1}$; \citealt{Perets2010, 2012ApJ...755..161K, 2017hsn..book..317T}). At later nebular epochs, they become ``rich'' in [Ca II] $\lambda\lambda$7291, 7324 emission lines with an integrated [Ca II]/[O I] $\lambda\lambda$6300, 6364 flux ratio $\ga 2$ \citep{Dan, Wynn}. 

Although Ca-rich transients may make up only a small fraction of all supernova events (10-20\% of SNe Ia; \citealt{De2020}, but see also \citealt{Frohmaier2018}), knowing their origin is nevertheless critical in understanding their role in chemical enrichment of the universe. Ca-rich transients may produce large amounts of calcium ($\sim 0.1$\,M$_{\odot}$; \citealt{Perets2010}), and in turn may be significant contributors to the calcium content of the intracluster medium \citep{2014ApJ...780L..34M}. However, more recent theoretical investigations \citep{DH15,Wynn,Polin21}, and analyses leveraging optical+near-infrared observations \citep{Dan}, have instead supported the notion that these explosions are far less abundant in calcium production than originally suspected (0.006 to 0.02\,M$_{\odot}$).

\begin{figure*}[ht!]
\begin{center}
\includegraphics[width=0.85\textwidth]{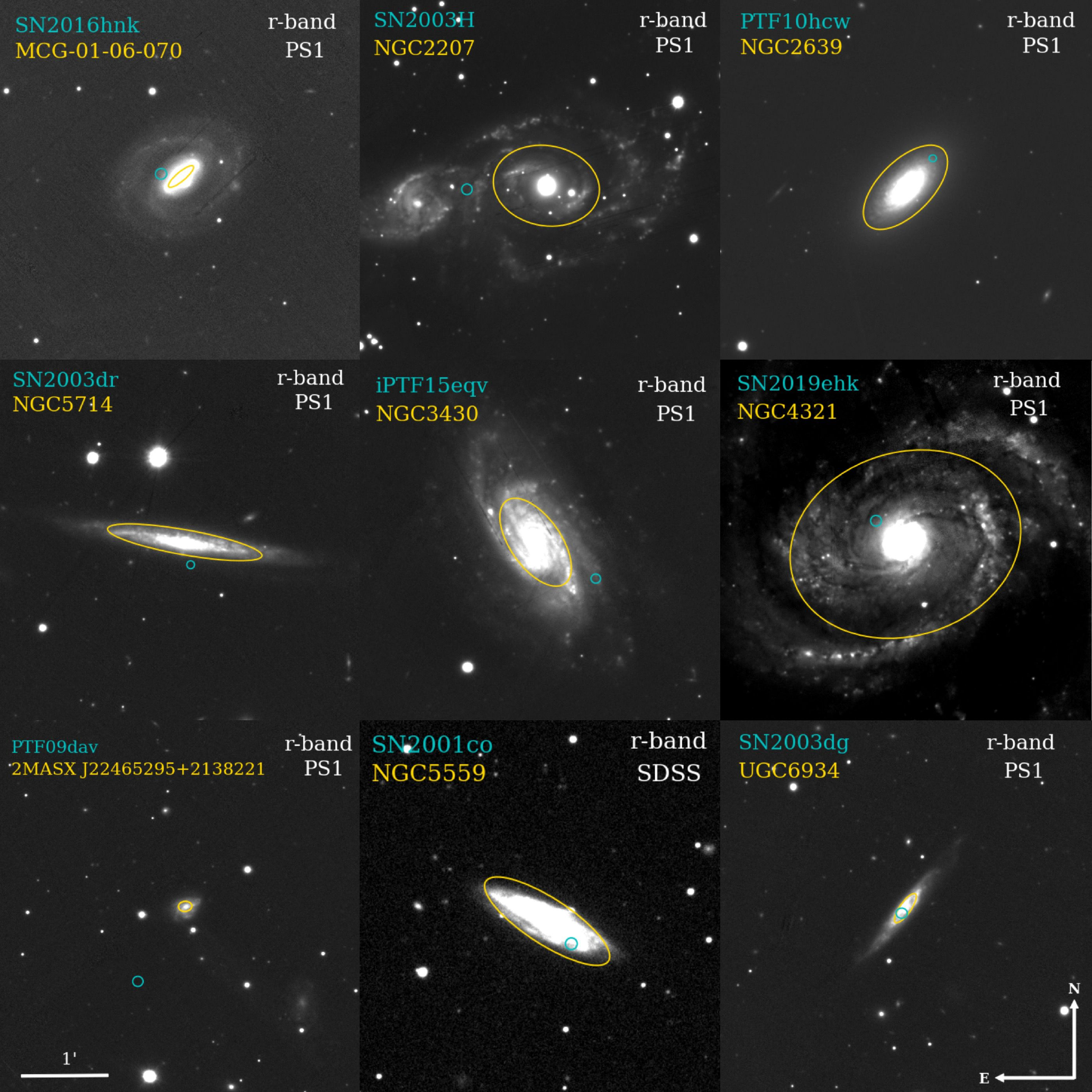}
\caption{Spiral host galaxies in our sample as observed by Pan-STARRS and SDSS. Each panel is $4 \times 4$ arcmin\(^2\) and aligned North up and East left with the same scale. The name of the supernova and host galaxy are indicated on the top left hand corner while the filter and survey are displayed on the top right hand corner. All host galaxies are circled in yellow and transients in blue. The ellipses are at twice the isophotal radius.}
\end{center}

\label{fig:panels-stype}
\end{figure*}

Many progenitor models have been suggested to explain the origin of Ca-rich transients. Ca-rich transients are often located at large offset distances away from their suspected host galaxies ($\sim$1/3 are offset $>20$ kpc; \citealt{2012ApJ...755..161K,2015MNRAS.452.2463F,Lyman2016}), and nearly half are discovered in older galaxies that exhibit sparse star formation, which together suggest a connection with white dwarf (WD) stars and an explosion potentially triggered by a helium detonation process \citep{Bildsten2007,SB09,Perets2010,WK2011}. However, the spectroscopic similarities between a large fraction of Ca-rich transients and core-collapse SNe which are found in galaxies with ongoing star formation, instead suggest connections with massive stars \citep{Kawabata2010,Dan,Lee2019}. Binary evolution involving a low-mass He or C$/$O progenitor star with a neutron star companion producing an ``ultra-stripped'' SN has also been suggested \citep{Tauris2015}, as have mergers between neutron stars and WDs \citep{Metzger2012,2014MNRAS.444.2157L}, and the tidal detonation of a WD by a neutron star or black hole \citep{Rosswog2008, Clausen2011, Metzger2012, MacLeod2014, Sell2015, Margalit2016, Bobrick2017, Zenati2019}. 

\begin{figure*}[ht!]
\begin{center}
\includegraphics[width=0.85\textwidth]{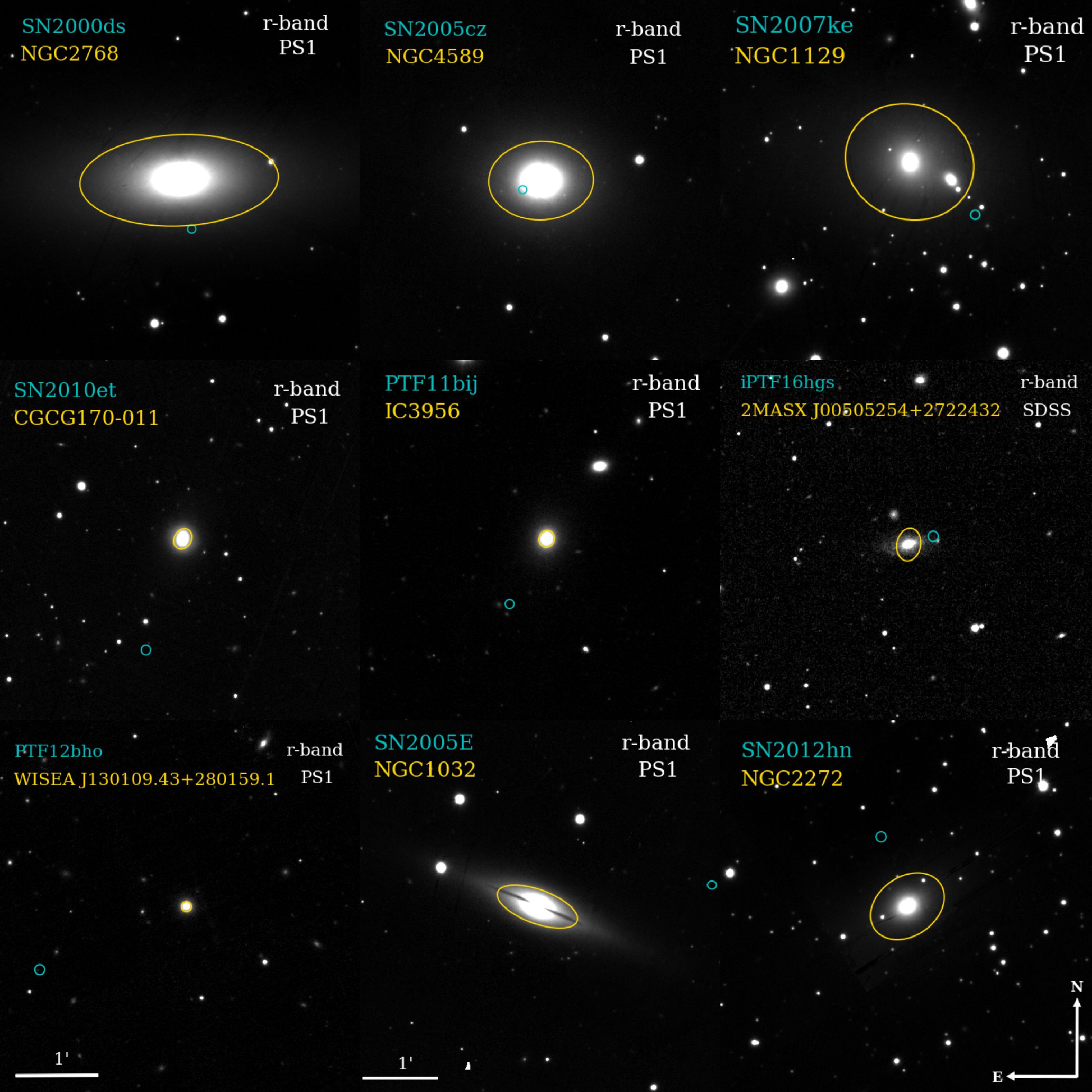}
\caption{Same as Figure~\ref{fig:panels-stype}, but for all elliptical host galaxies in our sample. Each panel is $4 \times 4$ arcmin\(^2\). This collage also includes NGC 1032, a S0/A type galaxy with a panel size of $5 \times 5$ arcmin\(^2\).}
\end{center}

\label{fig:panels-etype}
\end{figure*}

Recently, \cite{kish} distinguish Ca-rich transients into Ca Ib/c and Ca Ia sub-classes by features observed in photospheric phase spectra that are similar to Type Ia and Ib/c SNe, respectively.  In this system, Ca-Ia objects are believed to be a complete double detonation of low mass WDs where the efficiency of He burning in the outer ejecta is high, whereas Ca Ib/c objects are produced from a continuum of double detonations on even lower mass WDs (Ca-Ic) to He shell-only detonations or deflagrations (Ca-Ib) all with low He burning efficiency.

\begin{figure}[ht!]
\begin{center}
\vspace*{4.5mm}
\includegraphics[width=0.95\linewidth]{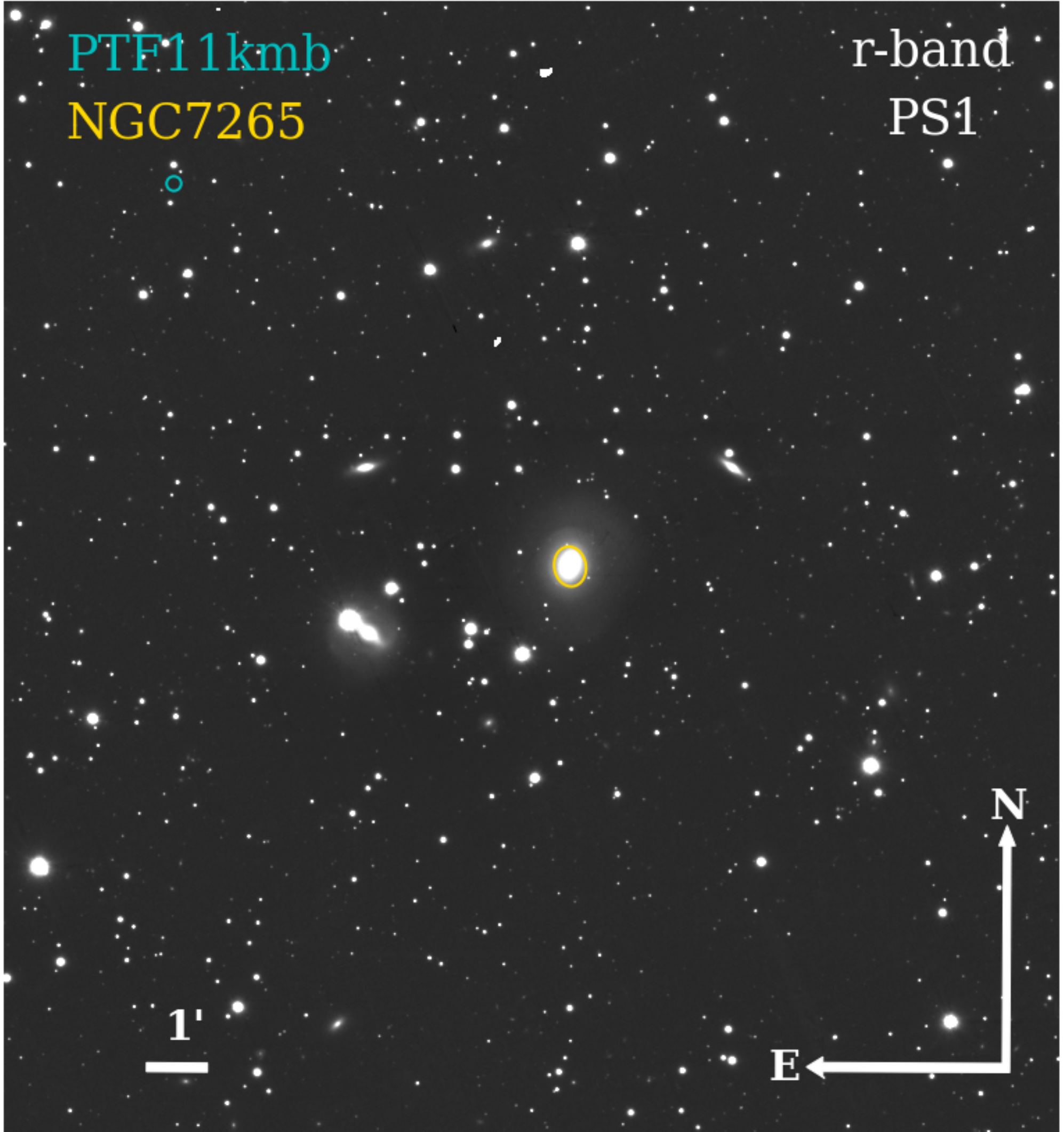}
\caption{Same as Figure~\ref{fig:panels-stype} and \ref{fig:panels-etype}, but for the suspected host galaxy of PTF11kmb. The panel is $15 \times 15$ arcmin\(^2\).}
\end{center}

\label{fig:PTF11kmb}
\end{figure}

The possibility that multiple progenitor channels contribute to the observed diversity of Ca-rich transients has not yet been ruled out, and the debated nature of many well-observed Ca-rich transients continues to fuel this uncertainty. Most recently, multiple progenitor scenarios have been suggested for SN\,2019ehk in NGC4321 (M100), the first and only Ca-rich transient with X-ray detections: an ultra-stripped SN \citep{Nakaoka}, a helium shell detonation on a sub-Chandrasekhar WD \citep{Wynn,Jacobson2021}, interaction of lowest mass massive star (9.5 -- 10 M$_{\odot}$) binaries \citep{Wynn}, and core collapse from a stripped-envelope massive star \citep{De2021}. Further complicating efforts to unify the class are a variety of objects with properties that bridge classifications. For example, the peculiar iPTF15eqv was more luminous and decayed much more slowly than other Ca-rich transients, but had one of the highest [Ca II]/[O I] flux ratios among all Ca-rich transients \citep{Dan}.  

To date, there has been considerable effort to constrain potential progenitor systems of Ca-rich transients using host galaxy information (see, e.g., \citealt{Yuan2013,Lunnan17,Shen2019, Perets2021}). Here we further these efforts by presenting a comprehensive analysis of the host galaxies of all reported and suspected Ca-rich transients using publicly available photometry and the parametric and nonparametric models in the advanced stellar inference Python package \textsc{prospector} \citep{Leja2017, Johnson21}.  We examine the properties of these host galaxies, including star formation, age, mass, and metallicity, to further constrain the potential progenitor systems of Ca-rich transients. The layout of the paper is structured with the following sections. Section \ref{sec:Observations} includes the various catalogs and bands used to create our data sets. Section \ref{sec:SED Modeling} describes the SED modeling using \textsc{prospector}. The results from using the parametric and nonparametric models are in Section \ref{sec:results}. Section \ref{sec:discussion} discusses and summarizes the interpretation and comparisons between our results in the two models.


\begin{deluxetable*}{ccccc}[t!]
\linespread{1.2}
\tablecaption{Ca-rich Transient Samples $\&$ Host Galaxies \label{tab:hostinfo}}
\tablecolumns{5}
\tablewidth{0pt}
\tablehead{
\colhead{SN} & 
\colhead{Host Galaxy} & 
\colhead{Host Type} &
\colhead{Host Redshift} & 
\colhead{Reference}
}

\startdata
2000ds & NGC 2768 & E6 & 0.00451 & 6 \\
2001co & NGC 5559 & SBb & 0.01723 & 6 \\
2003H & NGC 2207 & SABb/c & 0.00914 & 6 \\
2003dg & UGC 6934 & SAc/d & 0.01835 & 6 \\
2003dr & NGC 5714 & Scd & 0.00746 & 4, 6 \\
2005E & NGC 1032 & S0/A & 0.00899 & 1, 6 \\
2005cz & NGC 4589 & E2 & 0.00660 & 2, 6 \\
2007ke & NGC 1129 & E0 & 0.01733 & 6\\
2010et & CGCG 170-011 & E0 & 0.02334 & 6 \\
2012hn & NGC 2272 & E & 0.00710 & 5, 6 \\
2016hnk & MCG-01-06-070 & SAB & 0.01627 & 10, 11, 13 \\
2019ehk & NGC 4321 & SAB(s)bc & 0.00524 & 12, 14, 15 \\
PTF09dav & 2MASX J22465295+2138221 & S (Disturbed) & 0.03710 & 6, 3\\
PTF10hcw & NGC 2639 & SAa & 0.01113 & 8 \\
PTF11bij & IC 3956 & E3 & 0.03471 & 4, 6\\
PTF11kmb & NGC 7265 & S & 0.01696 & 8 \\
PTF12bho & WISEA J130109.43+280159.1 & E & 0.02339 & 8 \\
iPTF15eqv & NGC 3430 & SAB & 0.00529 & 7 \\
iPTF16hgs & 2MASX J00505254+2722432 & S & 0.01700 & 9 \\
\hline
\enddata
\tablecomments{Information on the Ca-rich samples included in this work and their putative host galaxies. All redshifts are from NED with the exception of the host of iPTF16hgs which comes from \cite{2018De}. Classifications of host galaxies are also from NED \citep{DeVau1964}. \\
References: (1): \citep{Perets2010}, (2): \cite{Kawabata2010}, (3): \cite{Sullivan2011}, (4): \cite{2012ApJ...755..161K}, (5): \cite{Valenti2014}, (6): \cite{2015MNRAS.452.2463F}, (7): \cite{Dan}, (8): \cite{Lunnan17}, (9): \cite{2018De}, (10): \cite{Sell2018}, (11): \cite{Galbany2019}, (12): \cite{Wynn}, (13): \cite{Jacobson2020}, (14): \cite{2021De}, (15): \cite{Jacobson2021}}
\end{deluxetable*}

\section{Observations} \label{sec:Observations}

The nineteen Ca-rich transients and their associated host galaxies found in the literature and used in this paper are listed in Table \ref{tab:hostinfo}. Images are shown in Figures \ref{fig:panels-stype}-\ref{fig:PTF11kmb}. The host morphology is determined according to classifications given by \cite{DeVau1964}. We utilize broadband photometry from the \textit{z = 0 Multi-wavelength Galaxy Synthesis I} (z0MGS; \citealt{2019ApJS..244...24L}), which is an atlas that contains ultraviolet (UV) and infrared (IR) images as observed by the Galaxy Evolution Explorer (GALEX; \citealt{Martin2005}) and Wide-field Infrared Survey Explorer (WISE; \citealt{2010AJ....140.1868W}) surveys. The atlas selects a sample of local galaxies (d $\lesssim$ 50 Mpc) from the Lyon Extragalactic Database (LEDA; \citealt{2003A&A...412...57P}; \citealt{2014A&A...570A..13M}) and measures integrated-light photometry in FUV, NUV, W1, W2, W3, and W4 bands. The maps contain matched resolution and astrometry to ensure consistency across all resolved measurements. 

We supplement these data with photometry from a variety of catalogs. Additional optical data are retrieved from the Sloan Digital Sky Survey Data Release 13 (SDSS DR13; \citealt{2017ApJS..233...25A}). SDSS DR 13 includes all coverage of prior data releases in the northern hemisphere and recalibrations of SDSS imaging catalogs. We correct for Galactic dust extinction for the optical photometry using Galactic reddening maps from \cite{1998ApJ...500..525S} and NASA/IPAC Extragalactic Database (NED).\footnote{\url{https://ned.ipac.caltech.edu/}} IR photometry in J-, H-, and K-band are extracted from the 2MASS All-Sky Extended Source Catalog (XSC; \citealt{2006AJ....131.1163S}) in the NASA/IPAC Infrared Science Archive (IRSA; \citealt{2006AJ....131.1163S}) and the 2MASS Large Galaxy Atlas (LGA; \citealt{Jarrett2003}) cataloged by \cite{Bai2015}. 

Photometry in $g$, $r$, and $z$ are included from the Legacy Surveys Data Release 8 (DR 8), as well as mid-IR photometry in the $\SI{3.4}{\micro\metre}$ and $\SI{4.6}{\micro\metre}$ WISE bands from the DESI Legacy Imaging Surveys \citep{2019AJ....157..168D}. While both the DESI survey and z0MGS atlas employ data from the WISE mission, the former targets the two shortest-wavelengths, W1 and W2, and measures the photometry using \textit{The Tractor} algorithms \citep{2016ascl.soft04008L} while the latter contains all four WISE bands directly from imaging. All photometry measurements in DESI are aperture-matched using \textit{The Tractor} package.

We also incorporate optical photometry from Pan-STARRS Data Release 2 (PS2; \citealt{2018AAS...23143601F}) and HyperLeda \citep{Paturel2003, Makarov2014} compiled by \cite{Bai2015} for galaxies that do not have any optical photometry from SDSS DR13 or DESI DR8. Using the equations and coefficients from \cite{2012ApJ...750...99T}, we transform $g$, $r$, $i$, and $z$ magnitudes in PS2 to SDSS bandpasses and keep the $y$ band as is. Since photometry from DESI and PS2 are not extinction-corrected, we perform the same Galactic dust extinction correction as before. For uniformity and consistency in the data, we convert all broadband photometry to AB magnitudes. In the case of three host galaxies, NGC 2207, UGC 6934, and MCG-01-06-070, photometry from 2MASS and DESI were inconsistent across other surveys, and were excluded from our fits. The mismatch is most likely the result of inconsistent aperture extraction and/or poorly determined zero-points. The complete photometric data utilized in our models are presented in the Appendix.

SED modeling relies heavily on the quality and completeness of photometry, and our host galaxy samples have a non-uniform number of data points. \cite{Johnson21} demonstrated that for a small number of photometric bands, the posteriors are largely determined by the prior distributions. Nevertheless, the addition of even a single data point can be informative given a reasonable choice of priors and Bayesian reasoning in high-dimension parameter spaces. An increase in the number of photometry bands will only allow the constraints on each parameter to become stronger and distinct from the prior distributions. In light of this, all photometric values are utilized in the SED modeling despite the unevenness of the data sets.

\section{SED Modeling} \label{sec:SED Modeling}

We model host galaxy SEDs with \textsc{prospector}, which is an advanced Python package that infers stellar population properties conditioned on the available photometric data \citep{Leja2017, Johnson21}.  We choose \textsc{prospector} as opposed to other similar implementations such as \textsc{magphys} \citep{2008MNRAS.388.1595D} and \textsc{cigale} \citep{2019A&A...622A.103B} to take advantage of its larger, more flexible parameter space. \textsc{prospector} also allows users to specify their priors on-the-fly and examine the resulting effect on their posteriors. Further, \textsc{prospector} fits the observational data using \texttt{dynesty} \citep{2020MNRAS.493.3132S}, a dynamic nested sampling method that is more efficient than the traditional Markov Chain Monte Carlo (MCMC) approach. Besides the spectroscopic redshift for the host of iPTF16hgs from \cite{2018De}, heliocentric redshift is adopted from NED in all photometric fits. For photometry with small observational uncertainties, a 5 \% error floor is put in place. This is to account for systematic model uncertainties as stellar population models are products of an amalgamation of theory, empirical libraries, and occasionally, ad-hoc assumptions \citep{Conroy2009, Conroy2013, Leja2019b, Alsing2020}.

To build the stellar population models, \textsc{prospector} utilizes the \texttt{Flexible Stellar Populations Synthesis} (\texttt{FSPS}) code \citep{Conroy2009, FSPS} and \texttt{Python-fsps} \citep{FSPSpython} for Python interface. It also uses WMAP9 cosmology \citep{Hinshaw2013} internally. For our work, we employ both a parametric and a more sophisticated nonparametric \texttt{continuity{\textunderscore}sfh} model to compare the results under different assumptions in each respective model. We use standard GALEX, Bessell, SDSS, Pan-STARRS, 2MASS, and WISE transmission curves in our photometry data that are fitted in the SED modeling. For consistency and a reasonable comparison, we also employ a Chabrier initial mass function (IMF) \citep{Chabrier2003} and Milky Way Extinction law \citep{1989ApJ...345..245C} in both models.

Dust plays a critical role in nearly all galaxies as it obscures light in the UV and emits it in the IR wavelengths \citep{Conroy2013}. In particular, dust emission from dust grains dominates the SED redwards of $>$ $\SI{3}{\micro\metre}$. The dust grains are postulated to be polycyclic aromatic hydrocarbon (PAH) particles as mixtures of amorphous silicate and graphite \citep{Draine&Li}. Thus, for any hosts that have photometry $>$ $\SI{3}{\micro\metre}$ (i.e. WISE bands), we implement the dust emission model from \cite{Draine&Li} to describe the PAH thermal emission features. The model is parametrized by the mass fraction of dust in PAH form, $qPAH$, the minimum radiation strength $U_{min}$, and the fraction $\gamma$ of dust in high radiation fields. These metrics are set as free parameters in both models to allow for more accurate uncertainties in the posterior distributions.

In both models, we incorporate nebular continuum in our model spectra and impose a 2:1 ratio on the amount of dust attenuation between the young and old stellar populations for the star-forming (SF) galaxies as younger stars tend to attenuate roughly twice the amount of dust in SF regions as older stars \citep{Calzetti2000}. The stellar mass -- stellar metallicity ($M_* - Z$) relationship from \cite{Gallazzi2005} is implemented as a prior for all host samples at their respective redshifts, following the implementation in \cite{Leja2017} where the width of the relationship is doubled to account for potential systemic errors. 


\begin{figure*}
\centering
\includegraphics[width=0.8\textwidth]{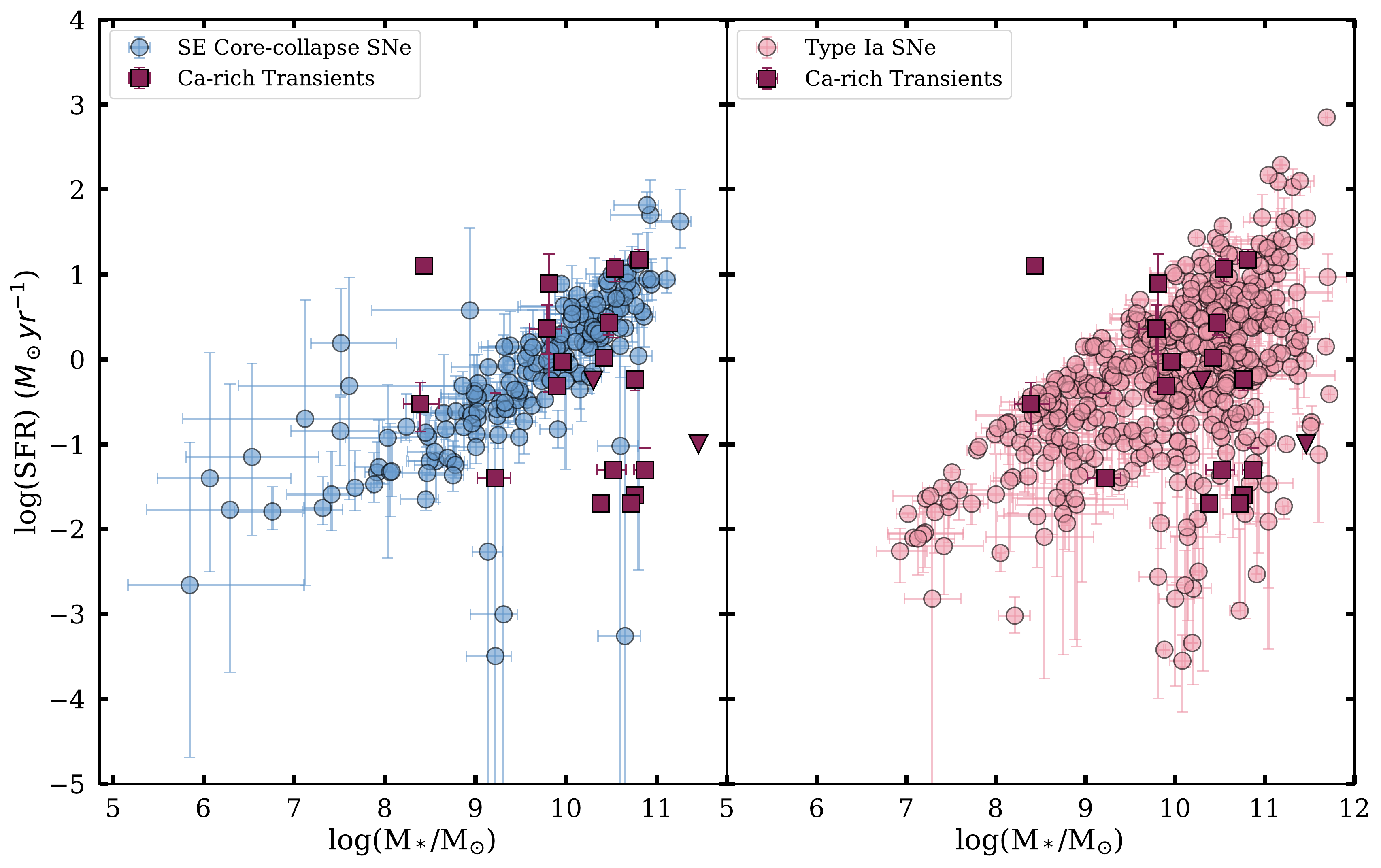}
\caption{Distribution of star formation rate as a function of stellar mass of all host galaxies on log scales. 3-$\sigma$ upper limits of SFR are displayed ($\blacktriangledown$) in the case of SN2007ke and SN2010et. Also shown are 167 SECC SNe host galaxies (blue; \citealt{Schulze2021}) and 442 Type Ia (pink; \citealt{Childress2013}) for comparison. Overall, the distribution of Ca-rich host galaxies displays no noticeable preference of galaxy populations for either type of SNe.}
\label{fig:sfr_age}
\end{figure*}

\subsection{Parametric Model} \label{sec:Parametric Model}
In our analyses, we first use a parametric model that is built with five unbiased free parameters: total mass in stars and remnants ($M$), metallicity ($Z$), star formation timescale ($\tau$), look-back time at which star formation commences (t$_{age}$), and dust extinction (A$_V$). In this model, the star formation history (SFH) is parametrized by the delayed-$\tau$ function form from \cite{2014arXiv1404.0402S}. Moreover, the star formation rate (SFR) is given by:

\begin{equation*}
    \text{SFR}(t) = M\times \Big[\int_0^t{t'e^{-t'/\tau} dt'}\Big]^{-1} \times t e^{-t/\tau},
\end{equation*}

\noindent{where $\tau$ is a free parameter that describes a star formation timescale for the exponentially declining SFH, $M$ is the total mass formed, and t=t$_{age}$ \citep{Nugent2020}.} 

We use t{$_{age}$} and $\tau$ to calculate $t_m$, the mass-weighted age through

\begin{equation*}
    t_m = t_{age} - \frac{\int_0^{t_{age}} t \times \text{SFR}(t) dt}{ \int_0^ {t_{age}} \text{SFR}(t) dt}
\end{equation*}

\noindent{\citep{Nugent2020} to compare ages in our samples. This is a more robust metric of age than t{$_{age}$}, which only describes the start of star formation and provides a better estimate compared to light-weighted age since it does not over-weigh the light from brighter and younger stars. The mass from the parametric fits denotes the total mass formed in the galaxies that includes the effects of stellar mass through stellar evolution such as SNe and Asymptotic Giant Branch (AGB) star winds. We convert total mass to stellar mass using $t_m$ and M at the time of the observation. We will only quote mass-weighted ages ($t_m$) and stellar masses ($M_*$) hereafter.}

\subsection{Nonparametric Continuity Model}

The \texttt{continuity\_sfh model} is one of five nonparametric priors available in \textsc{prospector}. In the parametric model, the SFH is constrained to a rigid function form; the nonparametric models are developed to allow for extra flexibility in SFHs that the parametric model is not attuned for. Specifically, \texttt{continuity\_sfh model} emphasizes a smooth SFH that is more sensitive to recent star bursts than the parametric model \citep{Leja2019} and offers a less biased reconstruction of the SFH solely based on the SED \citep{Conroy2013}.

For this model, we adopt a simple stellar population (SSP) to describe the SED temporal evolution of the stellar populations with a total of four parameters: metallicity (Z), dust ($A_V$), total mass formed (M), and SFR ratios between adjacent age bins. The total mass is split into N bins of equal mass m. We calculate a mass-weighted age for our samples using the following equation:

\begin{equation*}
    t_m = \frac{\sum\limits_{n=1}^{N} m_n \times age_n}{m_{total}}
\end{equation*}

\setlength{\tabcolsep}{6pt} 
\renewcommand{\arraystretch}{1.5} 
\begin{deluxetable*}{l|ccccccc}[!t]
\tablecolumns{8}
\tablewidth{0pt}
\tablecaption{Host Galaxies' Stellar Population Properties for the Parametric Model
\label{tab:parametric results}}
\tablehead{
\colhead {Transient}	 &
\colhead{$t_m$} &
\colhead {$\log(Z/Z_{\odot})$}  &
\colhead{log(M$_*$/M$_{\odot})$} &
\colhead {$\tau$}		    &
\colhead {$A_V$} &
\colhead{SFR$_{SED}$} & 
\colhead{log(sSFR)} \\
\colhead{} & 
\colhead{Gyr} & 
\colhead{} & 
\colhead{} & 
\colhead{} & 
\colhead{mag} & 
\colhead{$M_{\odot}$~yr$^{-1}$} & 
\colhead{yr$^{-1}$} 
}
\startdata
SN2000ds & $6.50^{+0.47}_{-0.38}$ & $-0.09 \pm 0.02$ & $10.76 \pm 0.02$ & $0.70^{+0.05}_{-0.04}$ &  $< 0.02$ & $0.025 \pm 0.001$ & $-12.37 \pm 0.03$ \\
SN2001co & $5.99^{+0.97}_{-0.11}$ & $0.14^{+0.03}_{-0.05}$ & $10.47 \pm 0.05$ & $3.27^{+1.13}_{-0.87}$ & $1.03^{+0.14}_{-0.16}$ & $2.68^{+0.83}_{-0.73}$ & $-10.04^{+0.16}_{-0.18}$ \\
SN2003H & $3.92^{+1.32}_{-1.39}$ & $0.15^{+0.03}_{-0.05}$ & $10.81^{+0.07}_{-0.11}$ & $5.44^{+2.28}_{-2.20}$ & $1.03^{+0.14}_{-0.16}$ & $14.97^{+4.78}_{-2.62}$ & $-9.94^{+0.24}_{-0.14}$ \\
SN2003dg & $2.27^{+2.13}_{-1.37}$ & $-0.22^{+0.21}_{-0.24}$ & $9.79^{+0.16}_{-0.19}$ & $2.84^{+3.44}_{-1.60}$ & $1.26^{+0.38}_{-0.37}$ & $2.31^{+2.06}_{-1.15}$ & $-9.43^{+0.48}_{-0.45}$ \\
SN2003dr & $7.35^{+0.34}_{-0.49}$ & $0.18 \pm 0.01$ & $9.90^{+0.02}_{-0.03}$ & $3.47^{+0.47}_{-0.30}$ & $0.52^{+0.09}_{-0.06}$ & $0.49^{+0.10}_{-0.06}$ & $-10.21^{+0.10}_{-0.07}$ \\
SN2005E & $6.90^{+2.38}_{-2.45}$ & $-0.34 \pm 0.10$ & $10.87^{+0.08}_{-0.12}$ & $0.80^{+0.37}_{-0.50}$ & $0.18^{+0.05}_{-0.06}$ & $0.05^{+0.04}_{-0.05}$ & $-12.47^{+0.19}_{-1.77}$ \\
SN2005cz & $5.83^{+1.02}_{-0.76}$ & $-0.14^{+0.05}_{-0.08}$ & $10.72^{+0.05}_{-0.04}$ & $0.61^{+0.11}_{-0.09}$ & $0.03 \pm 0.02$ & $0.021^{+0.002}_{-0.001}$ & $-12.40^{+0.05}_{-0.06}$ \\
SN2007ke & $10.37^{+1.56}_{-2.45}$ & $-0.09^{+0.13}_{-0.19}$ & $11.46^{+0.04}_{-0.05}$ & $0.31^{+0.33}_{-0.16}$ & $0.37 \pm 0.02$ & $< 0.10$ & $< -12.46$ \\
SN2010et & $5.12^{+1.53}_{-1.29}$ & $-0.66^{+0.12}_{-0.08}$ & $10.30^{+0.07}_{-0.08}$ & $0.33^{0.32}_{-0.17}$ & $0.09^{+0.08}_{-0.06}$ & $< 0.56$ & $< -10.55$ \\
SN2012hn & $4.69^{+1.02}_{-0.85}$ & $-0.21^{+0.09}_{-0.08}$ & $10.38^{+0.06}_{-0.05}$ & $0.54^{+0.12}_{-0.10}$ & $0.02 \pm 0.01$ & $0.023 \pm 0.001$ & $-12.01^{+0.06}_{-0.07}$\\
SN2016hnk & $7.94^{+0.25}_{-0.38}$ & $0.18^{+0.00}_{-0.01}$ & $10.42 \pm 0.02$ & $2.90^{+0.22}_{-0.16}$ & $0.17^{+0.16}_{-0.04}$ & $1.05^{+0.17}_{-0.11}$ & $-10.40^{+0.09}_{-0.06}$ \\
SN2019ehk & $0.89^{+0.96}_{-0.30}$ & $0.14^{+0.03}_{-0.06}$ & $10.54^{+0.06}_{-0.03}$ & $0.31^{+1.22}_{-0.14}$ & $0.78^{+0.07}_{-0.15}$ & $11.76^{+3.56}_{-3.52}$ & $-9.48^{+0.10}_{-0.17}$ \\
PTF09dav & $0.01^{+0.01}_{-0.00}$ & $0.18^{+0.01}_{-0.02}$ & $8.43^{+0.08}_{-0.05}$ & $1.21^{+3.80}_{-0.94}$ & $1.82^{+0.14}_{-0.10}$ & $12.70^{+3.05}_{-2.58}$ & $-7.18^{+0.13}_{-0.18}$\\
PTF10hcw & $3.03^{+0.85}_{-0.52}$ & $-0.26 \pm 0.09$ & $10.76^{+0.06}_{-0.05}$ & $0.49^{+0.14}_{-0.09}$ & $0.68^{+0.10}_{-0.15}$ & $0.58^{+0.11}_{-0.15}$ & $-11.01^{+0.10}_{-0.15}$ \\
PTF11bij & $5.42^{+3.04}_{-2.47}$ & $-0.22^{+0.20}_{-0.23}$ & $10.52^{+0.14}_{-0.18}$ & $0.58^{+1.23}_{-0.37}$ & $0.84^{+0.33}_{-0.30}$ & $0.05^{+1.74}_{-0.05}$ & $-11.78^{+1.63}_{-7.04}$ \\
PTF11kmb & $0.51^{+0.57}_{-0.20}$ & $-0.01^{+0.11}_{-0.26}$ & $9.81^{+0.07}_{-0.06}$ & $0.35^{+2.07}_{-0.21}$ & $2.53^{+0.35}_{-0.94}$ & $7.77^{+9.70}_{-7.36}$ & $-8.96^{+0.43}_{-1.25}$ \\
PTF12bho & $3.70^{+3.22}_{-2.11}$ & $-0.46^{+0.28}_{-0.21}$ & $9.22^{+0.17}_{-0.20}$ & $0.69^{+2.59}_{-0.48}$ & $0.46^{+0.57}_{-0.33}$ & $0.04^{+0.36}_{-0.04}$ & $-10.69^{+1.18}_{-4.68}$ \\
iPTF15eqv & $6.58^{+0.25}_{-0.37}$ &  $0.18 \pm 0.01$ & $9.96 \pm 0.02$ & $4.60^{+0.39}_{-0.42}$ & $0.26 \pm 0.02$ & $0.94 \pm 0.05$ & $-9.99 \pm 0.04$\\
iPTF16hgs & $0.82^{+1.49}_{-0.52}$ & $-0.26^{+0.24}_{-0.35}$ & $8.39^{+0.21}_{-0.18}$ & $3.17^{+3.87}_{-2.32}$ & $1.28^{+0.32}_{-0.44}$ & $0.30^{+0.23}_{-0.16}$ & $-8.90^{+0.43}_{-0.53}$ \\
\enddata
\tablecomments{Median values of the posterior distributions and 1-$\sigma$ uncertainties determined from the parametric model in \textsc{prospector}. All redshifts are helocentric. We fit for parameters t$_{age}$, M$_*$, A$_V$, $\tau$, and Z that describe stellar population properties of all host galaxies. We also derived t$_m$, SFR$_{SED}$ and sSFR using equations from Section \ref{sec:Parametric Model}, along with 3$\sigma$ upper limits on $A_V$ for the host of SN2000ds and SFRs for the hosts of SN2007ke and SN2010et.}
\end{deluxetable*}

\noindent{where $m_n$ is the mass formed in each bin, $age_n$ is the average age in each bin, and $m_{total}$ is the sum of masses in all age bins.  We also convert the total mass formed to stellar mass ($M_*$) using the same method as the parametric model and will refer to $t_m$ and $M_*$ for the rest of the paper.}

The continuity prior fits for $\Delta$log(SFR) between adjacent age bins with (N-1) SFH variable. We use eight bins for this analysis as results show that SED fits become largely insensitive to the number of temporal bins as long they surpass four bins \citep{Leja2019}. The first two bins are fixed at 0 -- 30 Myr and 30 -- 100 Myr with the youngest bin being the smallest to permit a maximally old stellar population \citep{Leja2019}. The upper limit for the age bins is calculated as the age of the Universe at the respective z of each host galaxy. All temporal bins are spaced equally in logarithmic time except the first two and the last bin since the separation is approximately proportional to the distinguishability of SSPs \citep{Ocvirk2006, Leja2019}.

Due to the flexibility of the \texttt{continuity\_sfh model}, the results are typically more reliable compared to the parametric fits. However, this extra parameter space is also more computationally expensive and most useful when it is constrained by high-quality data. Hence, we only apply this technique on nine host galaxies that are constraining enough to provide an accurate nonparametric SFH. The SED will encompass the observed photometry, observed photometry error, the model photometry, and model spectrum that depicts how well the model has been fit to the data.

\section{Results} \label{sec:results}

\subsection{Host Properties in Parametric Model}
The results from the parametric fit are summarized in Table \ref{tab:parametric results}. The mass ($M$) and metallicity ($Z$) parameters are on log scales. For the analysis, we regroup galaxies into more general classes of Elliptical (E-) and Spiral (S-) type. The hosts in E-type group exhibit an average $t_m$ of 5.95 Gyr and a median $t_m$ of 5.42 Gyr, implying relatively older stellar populations with less star formation activities. Conversely, the average $t_m$ is 3.85 Gyr and the median $t_m$ is 3.48 Gyr for S-type galaxies. S-type galaxies exhibit an overall younger stellar population, though the range spans from 0.01 to 7.94 Gyr.

As mentioned in Section \ref{sec:Parametric Model}, the parametric SFR is described by a delayed exponentially declining function. Here, a SFR-$M_{*}$ distribution is shown in Figure \ref{fig:sfr_age}. The data points for the host of SN2007ke and SN2010et are different from others to distinguish upper limits on the SFR. For comparison, we also show the same metrics for Type Ia and stripped-envelope CC (SECC; Type IIb, Ib/c) SNe host galaxies from \cite{Childress2013} and \cite{Schulze2021}, respectively. Similar to our analysis, the authors calculated SFR and $M_{*}$ using photometry, the \cite{Chabrier2003} IMF, and an exponentially-declining SFH in both works. We find that the distribution of Ca-rich host galaxies generally follows those of Type Ia and SECC SNe hosts, and tends to favor more massive hosts with lower SFRs. No clear preference between the host distributions of Type Ia and SECC SNe is observed. Notably, samples with higher stellar masses (log(M$_*$/M$_{\odot})$ $>$ 10.5) and lower metallicities ($\log(Z/Z_{\odot})$ $<$ $\sim$ -0.10) have low SFRs and are predominately E-type galaxies. Models from \cite{Yates2012} suggests that the galaxies may have undergone gradual dilution of gas phases after merger events, thereby halting further star formation. 

Also displayed in Figure \ref{fig:sfr_age} is the host of PTF09dav as an outlier in the distribution. Such an offset from the distribution is in contrast with the fact that PTF09dav was classified to be a peculiar subluminous Type Ia SN \citep{Sullivan2011}. The galaxy has a disturbed morphology and nebular emission lines in the spectrum that are indicative of ongoing star formation \citep{Sullivan2011}, consistent with its high SFR and more aligned with SECC SNe host distribution than Type Ia. 

In the case of iPTF16hgs in which the host is a low-metallicity, star-forming, spiral dwarf galaxy, \cite{2018De} found a stellar mass log(M$_*$/M$_{\odot})$ = $8.81 \pm 0.02$, a mean stellar population age of $0.2^{+0.63}_{-0.02}$ Gyr, and an integrated SFR of 0.5 $M_{\odot}$~yr$^{-1}$. We compare our mass, metallicity, and stellar population age to that of \cite{2018De} and they are overall in good agreement with one another. The stellar mass from our parametric fit is slightly smaller at $\log(M_{\ast}/M_{\odot})$ = 8.39. However, parametric models are known to systematically underestimate stellar masses due to an offset from the true shape of the galaxy's SFH \citep{Carnall2019}. The mean $t_m$ from our analysis is 0.82 Gyr, just within the 1-$\sigma$ uncertainty from \cite{2018De}. The significantly sub-solar metallicity of $\approx$ 0.4 Z$_{\odot}$ ($\log(Z/Z_{\odot}$ = -0.38) reported by \cite{2018De} also agrees with our result $\log(Z/Z_{\odot})$ = -0.26 within 1-$\sigma$ uncertainty. Although the SFR from \cite{2018De} is estimated from H$\alpha$ maps, we note that it is still consistent with our SFR derived from the parametric fit. The differences in our models and observational data compared to what was applied by \cite{2018De} can also explain the discrepancy between our results.

\setlength{\tabcolsep}{6pt} 
\renewcommand{\arraystretch}{1.5}
\begin{deluxetable*}{l|ccccccc}[!t]
\tablecolumns{7}
\tablewidth{0pc}
\tablecaption{Host Galaxies' Stellar Population Properties for the Continuity Model
\label{tab:np results}}
\tablehead{
\colhead {Transient}	 &
\colhead {Host Galaxy}	 &
\colhead{$t_m$} &
\colhead {$\log(Z/Z_{\odot})$}  &
\colhead{log(M$_*$/M$_{\odot})$} &
\colhead {$A_V$} &
\colhead{log(SFR$_{SED}$)} \\
\colhead{} &
\colhead{} &
\colhead{Gyr} & 
\colhead{} & 
\colhead{} & 
\colhead{mag} & 
\colhead{$M_{\odot}$~yr$^{-1}$} & 
}

\startdata
SN2000ds & NGC 2768 & $8.77^{+0.42}_{-0.28}$ & $-0.06^{+0.02}_{-0.02}$ & $10.92 \pm 0.01$ & $<0.004$ & $-1.82^{+0.10}_{-0.27}$ \\
SN2001co & NGC 5559 & $7.77^{+1.06}_{-0.77}$ & $0.06^{+0.02}_{-0.04}$ & $10.66 \pm 0.04 $ & $0.90^{+0.15}_{-0.14}$ & $0.34^{+0.16}_{-0.17}$ \\
SN2003dr & NGC 5714 & $9.96^{+1.26}_{-2.32}$ & $0.06^{+0.03}_{-0.02}$ & $10.14 \pm 0.05$ &  $0.53^{+0.18}_{-0.02}$ & $-0.12^{+0.07}_{-0.03}$ \\ 
SN2005cz & NGC 4589 & $9.22^{+0.56}_{-0.32}$ & $-0.23 \pm 0.03$ & $10.94 \pm 0.01$ & $0.02^{+0.00}_{-0.01}$ & $-2.05^{+0.19}_{-0.65}$ \\
SN2012hn & NGC 2272 & $8.26^{+0.37}_{-0.44}$ & $-0.02^{+0.07}_{-0.08}$ & $10.58 \pm 0.02$ & $0.01^{+0.02}_{-0.01}$ & $-2.05^{+0.25}_{-0.48}$ \\
SN2016hnk & MCG-01-06-070 & $9.85^{+0.99}_{-1.93}$ & $0.09 \pm 0.01$ & $10.68 \pm 0.02$ &  $0.38^{+0.02}_{-0.04}$ & $0.33^{+0.03}_{-0.04}$ \\ 
SN2019ehk & NGC 4321 & $4.83^{+1.63}_{-1.53}$ & $0.06^{+0.03}_{-0.05}$ & $10.81^{+0.07}_{-0.08}$ &  $0.56^{+0.07}_{-0.05}$ & $0.73 \pm 0.18$ \\ 
PTF10hcw & NGC 2639 & $7.47^{+0.82}_{-0.57}$ & $-0.15^{+0.07}_{-0.10}$ & $11.05 \pm 0.03$ &  $0.40^{+0.13}_{-0.05}$ & $-0.77^{+0.20}_{-0.41}$ \\ 
iPTF15eqv & NGC 3430 & $8.31^{+0.65}_{-0.77}$ & $0.06^{+0.03}_{-0.05}$ & $10.21^{+0.02}_{-0.03}$ &  $0.35^{+0.08}_{-0.05}$ & $0.20^{+0.12}_{-0.07}$ \\ 
\enddata
\tablecomments{Median values and 1-$\sigma$ uncertainties of the posterior distributions using the \texttt{continuity\_sfh model}. A 3-$\sigma$ upper limit on $A_V$ is derived for NGC 2768. Similar to the parametric model, we fit for M$_*$,  A$_V$, and Z. But unlike the function form imposed on the SFH by the parametric model, we use a time binning method in the nonparametric model, allowing greater flexibility to reproduce more accurate SFHs. We determine the SFR and t$_m$ from the SFR ratios between adjacent time bins.}
\end{deluxetable*}

\begin{figure}
\centering
\includegraphics[width=0.48\textwidth]{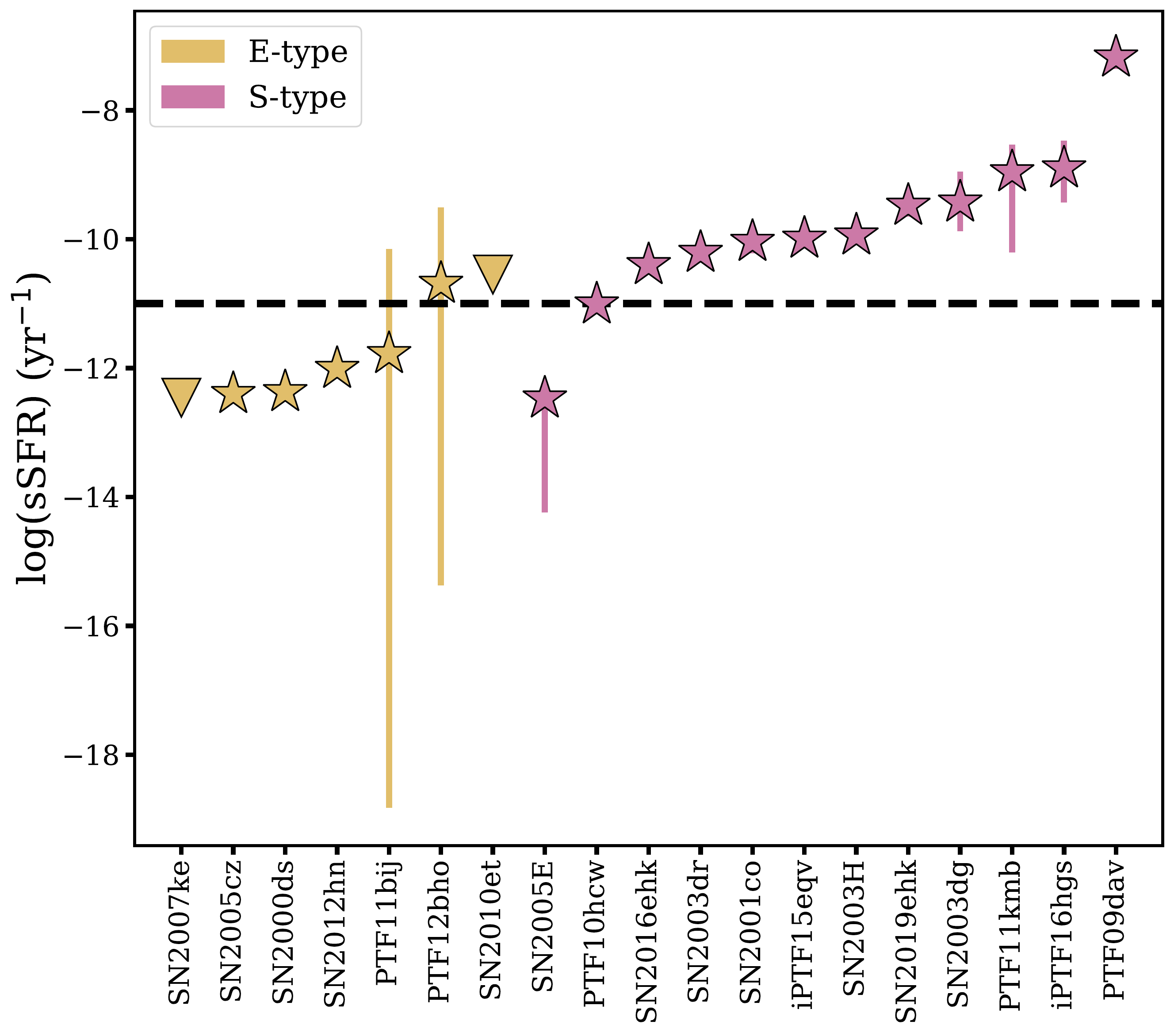}
\caption{sSFR of all host galaxies with Ca-rich transients labeled on the x-axis. The \textit{red and dead} threshold defined by \cite{Fontana} (sSFR $<$ -11 $yr^{-1}$) is shown as a black dashed line. E-type is marked in gold while S-type is marked in dark pink. The hosts of SN2005E and PTF10hcw are both disc galaxies that satisfy as \textit{red and dead}. The log(sSFR) for SN2007ke and SN2010et host galaxies are upper limits ($\blacktriangledown$).}   
\label{fig:ssfr}
\end{figure}


The metallicities of all Ca-rich transient host galaxies range between $\log(Z/Z_{\odot})$ = -0.66 and $\log(Z/Z_{\odot})$ = 0.18. In addition, we obtain an average stellar mass log(M$_*$/M$_{\odot})$ = 10.62, with the a minimum mass of log(M$_*$/M$_{\odot})$ = 8.28 and a maximum of log(M$_*$/M$_{\odot})$ = 11.46. A subset (10) of our Ca-rich transient host galaxies were previously analyzed by \cite{Yuan2013}. The authors converted observed K-band total luminosities of each host galaxy to stellar masses using numerical models of cosmic chemical evolution from \cite{Kobayashi2007} and found a range between log(M$_*$/M$_{\odot})$ = 10.08 and 11.15 (1.2 $\times$ $10^{10}$ $M_{\odot}$ -- 1.4 $\times$ $10^{11}$ $M_{\odot}$). While most of our results fall within the range, the hosts of PTF09dav, SN2003dg and SN2003dr are noticeably below the lower limit. Again, discrepancies in stellar mass likely arise from differences in methodologies and underlying assumptions.

In addition to the SFR, we also find the specific SFR (sSFR) defined as SFR per unit stellar mass for all host galaxies. We present the results in Figure \ref{fig:ssfr}. Galaxies at low redshifts are considered to be quiescent if their sSFR is $ <10^{-11}$\,yr$^{-1}$ \citep{Fontana}. Based on this quantitative threshold, we find that the majority of the E-type host galaxies have sSFR $<$ $10^{-11}$\,yr$^{-1}$, labeling them as ``red and dead.'' The host of PTF12bho (gold) is an exception to this trend. Similarly, the hosts of 2005E and PTF10hcw, both disc galaxies, also do not obey the expected pattern as S-type galaxies. NGC 1032 (SN2005E) is an S0/a galaxy that is in the transition period between E and Sa galaxies on the stage sequence, while NGC 2639 (PTF10hcw) is an SAa galaxy. These two galaxies produce large bulge light and are considered early-type systems along the Hubble sequence with strongly decreasing SFR, suggesting a dearth of new stars within the galaxies, resulting in a small sSFR \citep{Sandage1986}. \cite{Zhou2021} also found that massive red spirals such as NGC 2639 and NGC 1032 and massive elliptical galaxies have very similar SFHs. This means that they have only experienced one major star formation episode early on (and possibly fast quenching). Overall, more than $50\%$ of the host galaxies we examined exhibit some level of recent star formation that could support young star clusters.

\subsection{Host Properties with a Nonparametric Model}
We impose a nonparametric form of SFH to model the host galaxies of SN2000ds, SN2001co, SN2003dr, SN2005cz, SN2012hn, SN2016hnk, SN2019ehk, PTF10hcw, and iPTF15eqv. The sample contains three quiescent E-type galaxies (NGC 2768, NGC 4589, and NGC 2272) and six S-type galaxies (NGC5559, NGC5714, MCG-01-06-070, NGC4321, NGC2639, and NGC3430). Each of these host galaxies has at least 9 photometric values, which we found to be the minimum required for a confident analysis. We examine our results using the nonparametric model with various degrees of star formation activities and evaluate the effects on the stellar population properties. Parameters such as $M_*$ and $t_m$ are compared against metrics derived using the parametric model from Section \ref{sec:Parametric Model}. 

\begin{figure*}
\centering
\includegraphics[width=0.9\textwidth]{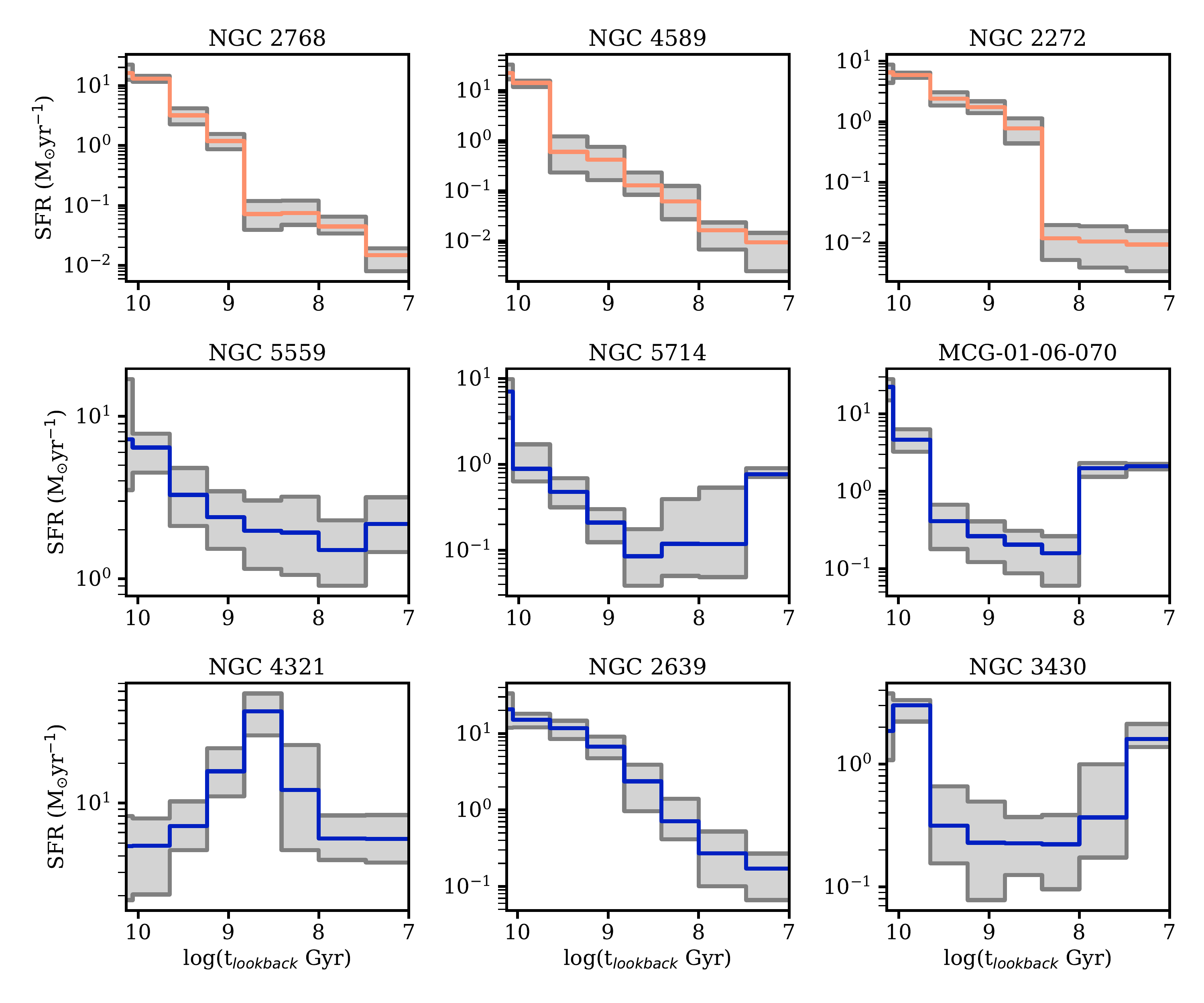}
\caption{The SFH of nine host galaxies constructed using the nonparametric \texttt{continuity\_sfh model}. The quiescent galaxies (NGC 2768, NGC 4589, and NGC 2272) have constant declines in the SFR with no recent burst of star formation. This contrasts with the star-forming galaxies, which shows evidence for a recent increase in SFR. One exception is NGC 2639 which resembles the steady decrease in SFRs found for E-type galaxies, and is consistent with our finding that it has the lowest SFR compared to other S-type galaxies in our sample.
\label{fig:sfh}}
\end{figure*}

The results of the nine host galaxy samples are shown in Table \ref{tab:np results}. The nine galaxies have an average $t_m$ of 8.27 Gyr, pointing to old stellar populations in the host environment. We find all three of the E-type galaxies have sub-solar metallicities with NGC 4589 being the most metal-poor. Conversely, almost all S-type galaxies have roughly the same super-solar metallicity. The nonparametric fits also show that the stellar masses of the four hosts are in the higher-mass range with the least massive being log(M$_*$/M$_{\odot})$ = 10.14 (NGC 5714) and the most massive being log(M$_*$/M$_{\odot})$ = 11.05 (NGC 2639) which are both S-type galaxies. 

We find that S-type galaxies are much dustier, averaging $A_V$ = 0.52 mag, unlike the quiescent galaxies in our sample. The dust extinction is an indicator of emission from young, massive stars and a higher SFR. Indeed, comparing log(SFR) between E-type and S-type galaxies, we find a remarkably higher log(SFR) in all S-type galaxies from the last 0 -- 30 Myr as opposed to the three quiescent hosts. The recent and ongoing star formation supports the possibility of a massive star progenitor behind the Ca-rich transient via core-collapse. However, the quiescent galaxies in our sample with low SFRs in the recent years challenge applying this interpretation for all Ca-rich transients. Notably, \cite{Crocker2011} determined a SFR range for NGC 2768 between 0.025 and 0.195 $M_\odot$~yr$^{-1}$,  which is consistent with our findings.

\begin{figure*}
\centering
\hspace*{-1.8cm}
\includegraphics[width=1.2\textwidth]{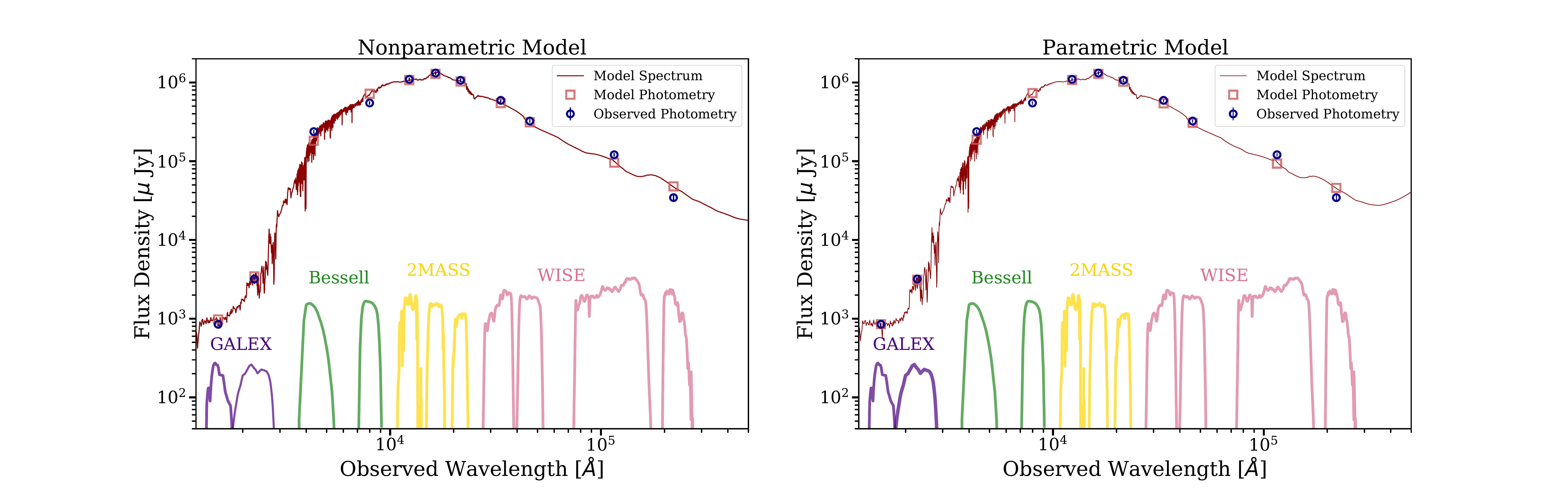}
\caption{The broad-band photometry of the host of SN2000ds ranging from optical to FIR (blue data points), and best-fit model spectrum and model photometry (red line and square data points) from the nonparametric (left) and parametric model fitting (right). There is an excellent consistency in the observed optical photometry, model photometry and features. GALEX UV-band, Bessell BI-band, 2MASS JHK-band, and WISE W1W2W3W4-band filter response curves are used in both fits.} 
\label{fig:2768_panel}
\end{figure*}

To further evaluate our results from the nonparametric fits, we construct a panel of SFHs with SFR as a function of lookback time using eight temporal bins as shown in Figure \ref{fig:sfh}. We find that the SFRs of E-type, quiescent galaxies, show steady exponential declines with very low present-day star formation rates. These SFHs provide evidence that the majority of stellar mass has been formed during the earlier epoch and only a small fraction of the mass is contributed by the recent star formation activities. Since a large percentage of mass was formed in the earlier temporal bins, we find consistency in the $t_m$ as most of the weight is determined by the older stellar population in the host galaxies. The SFH of four S-type galaxies also show a gradual decline in the galaxy's history. But unlike the E-type galaxies, the SFRs became roughly constant and displayed a slight increase $<$ 1 Gyr ago in lookback time. The climb in SFRs represents an association with new births of massive young stars within the host environment. NGC 2639 is a type SAa, Seyfert galaxy with LINER nuclear activity \citep{Lunnan, Biny19}. Its SFH is more similar to E-type galaxies with an apparent exponential decline and decreasing SFRs. This agrees with the result that NGC 2639 has the lowest recent SFR out of all S-type galaxy samples. \cite{Biny19} estimated a SFR on the order of 1 $M_{\odot}$~yr$^{-1}$ from {\it Spitzer} infrared fluxes, which is consistent with the low SFR found in the nonparametric fit. The morphology of the galaxy could indicate a past gas-rich merger that caused the drop in SFR \citep{Biny19}. 

Stellar population analysis was performed by \cite{Galbany2019} on the host galaxy of SN2016hnk using integral field spectroscopy. The authors report a stellar mass log(M$_*$/M$_{\odot})$ = 10.68, which is equivalent to our result (log(M$_*$/M$_{\odot})$ = 10.68). We applied a -0.24 dex correction \citep{Mitchell2013} to the stellar mass since a Chabrier IMF is used in this work as opposed to a Salpeter IMF \citep{Salpeter1955}. The corrected stellar mass for the host of SN2016hnk is still comparable to our results. \cite{Galbany2019} also derived an average stellar age = 690 Myr, SFR = 0.65 $M_{\odot}$~yr$^{-1}$ and A$_V$ = 0.02 mag for MCG-01-06-070. The reported age is light-weighted and cannot be directly compared with the mass-weighted age in our analysis, as light-weighed ages are always younger than mass-weighted determinations. Using the nonparametric \texttt{continuity\_sfh model}, we find SN2016hnk's host galaxy to be dustier and more star-forming in comparison. Broadly, the discrepancies are likely due to differing models of SFH and underlying assumptions. Although \cite{Galbany2019} employed a nonparametric SFH, they utilized STARLIGHT \citep{Fernandes2005, Fernandez2016} on the integrated spectrum of host galaxy to estimate the fractional contribution of different SSP instead of fitting photometric data to a nonparametric model.

\section{Discussion} \label{sec:discussion}

\subsection{Comparing Parametric and Nonparametric Results}

The parametric and nonparametric \texttt{continuity\_sfh model} are two distinctly different templates with varying stellar population metrics. The most notable difference is the fact that parametric SFH assumes a delayed $\tau$ function form for SFR(t) whereas the nonparametric model does not. The use of two different models allows us to examine the influence of our choice of priors and observational data on the inferred properties. So, it is pertinent that we compare the results from each model to extract more information from the SED fitting.

We first consider the effects on the mass-weighted age. \cite{Carnall2019} demonstrated that the parametric model underestimates stellar population age and favors galaxy formation at later times. Comparing $t_m$ of all nine host galaxies in each model, we find that the mass-weighted stellar ages obtained from the \texttt{continuity\_sfh model} are consistently older than what was derived using the parametric model, in agreement with the finding that parametric models systematically underestimate galaxy ages \citep{Wuyts2011}. The stellar population was found to be older in the nonparametric model by at least 1.5 Gyr with the largest offset being 4.44 Gyr, more than double the mass-weighted age from the parametric results. The fact that the parametric model struggles to obtain a better estimate for $t_m$ is due to its strong dependency on priors and the priors tendency to prefer recent star formation despite a more predominant contribution from the older stars in the overall stellar masses \citep{Lower2020}. Additionally, the \texttt{continuity\_sfh model} produces better stellar population ages when fit to simulations, outperforming the parametric model by producing a more realistic age for the host environments \citep{Lower2020}.

Stellar mass is generally considered the most robust metric derived from SED fitting. Nonparametric SFHs are found to recover stellar masses with higher accuracy attributable to its flexibility \citep{Lower2020}. The parametric model, on the other hand, underestimates stellar masses due to difficulty in fine-tuning the prior for SFR. Analysis from \cite{Carnall2019} reported a $\sim$ 0.3 dex uncertainties for stellar mass property while \cite{Lower2020} found an average offset in stellar mass from the parametric model to be 0.38 dex with an average uncertainty of 0.19 dex when compared to simulated galaxies. Our results show that both the E-type and S-type galaxies from our sample have a lower $M_*$ with comparable uncertainty range in the parametric model compared to the \texttt{continuity\_sfh model}, consistent with the fact that parametric models tend to underestimate stellar masses \citep{Carnall2019, Lower2020}. This also agrees with the finding that nonparametric SFH models infer an older stellar population age, which, in turn, produce a higher stellar mass \citep{Leja2019b}. Additionally, the stellar mass between the two models from our sample differ by an average of 0.23 dex and this bias comes from differences in the assumed SFH models.

We now consider the SFRs inferred in our sample. It is important to differentiate that the SFR found in the parametric model is derived from the overall stellar population age, while the nonparametric SFRs reported are only within the youngest age bin, i.e. 0 -- 30 Myr in lookback time. We find that the parametric model routinely offset from the nonparametric SFH model, ranging between 0.1 and 0.55 dex. This places the bias within the systematic uncertainties of $\sim$ 0.5 dex typically assumed for SFR measurements \citep{Pacifici2015}. The parametric model predicted higher SFRs for six out of nine hosts including all three quiescent galaxies. This is related to the model's strong bias by the SFH priors compared to other metrics such as stellar mass and dependence on the SFH over more recent years in the galaxy histories \citep{Carnall2019}. Nonetheless, both models infer low SFR(t) for the quiescent, E-type galaxies and moderate SFR(t) for the star-forming, S-type galaxy.

The evolution of SFRs with respect to the lookback times as shown in Figure \ref{fig:sfh} demonstrates that an exponentially declining function such as the delayed-$\tau$ model is, in fact, suitable for quiescent galaxies with very old ages. However, the shortcoming of applying the delayed-$\tau$ SFH model to galaxies is the false constraints imposed on galaxy properties by the SFH priors. In this case, the nonparametric model is a more promising alternative that provides greater flexibility than parametric models, and allow more direct incorporation of prior beliefs \citep{Lower2020}. As a consequence, it produces a more accurate SFH, especially with more bursty SFHs. 

\setlength{\tabcolsep}{6pt} 
\renewcommand{\arraystretch}{1.5} 
\begin{deluxetable*}{l|ccccc}[!t]
\tablecolumns{5}
\tablewidth{0pc}
\centering
\caption{Host galaxy SN Rates and Probabilities of Core-Collapse vs.\ Type Ia Origin.}
\tablehead{
\colhead {Transient}	 &
\colhead {Host Galaxy} &
\colhead{CC SN/yr} &
\colhead {Ia SN/yr}  &
\colhead{p$_{CC}$} &
\colhead{p$_{Ia}$}
}
\label{tab:SN Rates}
\startdata
SN2000ds & NGC 2768 & $(3.6_{-0.7}^{+1.0}) \times 10^{-4}$ & $(7.9_{-0.8}^{+0.9}) \times 10^{-3}$ & $0.04_{-0.01}^{+0.01}$ & $0.96_{-0.01}^{+0.01}$\\
SN2001co & NGC 5559 & $(2.7_{-0.6}^{+0.8}) \times 10^{-2}$ & $(7.8_{-1.4}^{+1.7}) \times 10^{-3}$ & $0.77_{-0.06}^{+0.05}$ & $0.23_{-0.05}^{+0.06}$ \\
SN2003dr & NGC 5714 & $(7.0_{-0.5}^{+0.7}) \times 10^{-3}$ & $(1.3_{-0.3}^{+0.5}) \times 10^{-3}$ & $0.85_{-0.05}^{+0.03}$ &  $0.15_{-0.03}^{+0.05}$\\
SN2005cz & NGC 4589 & $(2.4_{-1.1}^{+2.3}) \times 10^{-4}$ & $(7.5_{-1.1}^{+1.3}) \times 10^{-3}$ & $0.03_{-0.01}^{+0.03}$ &  $0.97_{-0.03}^{+0.01}$\\
SN2012hn & NGC 2272 & $(1.7_{-0.7}^{+1.5}) \times 10^{-4}$ & $(4.6_{-0.4}^{+0.5}) \times 10^{-3}$ & $0.03_{-0.02}^{+0.03}$ &  $0.97_{-0.03}^{+0.02}$\\
SN2016hnk & MCG-01-06-070 & $(2.6_{-0.2}^{+0.3}) \times 10^{-2}$ & $(4.6_{-0.8}^{+1.1}) \times 10^{-3}$ & $0.85_{-0.03}^{+0.02}$ & $0.15_{-0.02}^{+0.03}$ \\ 
SN2019ehk & NGC 4321 & $(7.8_{-1.7}^{+2.3}) \times 10^{-2}$ & $(4.1_{-1.0}^{+1.4}) \times 10^{-2}$ & $0.65_{-0.07}^{+0.07}$ &  $0.35_{-0.07}^{+0.07}$\\
PTF10hcw & NGC 2639 & $(3.6_{-1.5}^{+2.8}) \times 10^{-3}$ & $(1.6_{-0.2}^{+0.3}) \times 10^{-2}$ & $0.18_{-0.07}^{+0.1}$ &  $0.82_{-0.1}^{+0.07}$\\
iPTF15eqv & NGC 3430 & $(1.5_{-0.2}^{+0.2}) \times 10^{-2}$ & $(2.0_{-0.4}^{+0.5}) \times 10^{-3}$ & $0.89_{-0.03}^{+0.02}$ &  $0.11_{-0.02}^{+0.03}$\\
 \enddata
\end{deluxetable*}

To illustrate the quality of our model fits, we display the SED of NGC 2768 using each model in Figure \ref{fig:2768_panel}. The SED coverage spans from UV (GALEX) to FIR (WISE), representing the full broadband dataset which impacts the accuracy of the galaxy properties inferred from SED fitting. The shape of the SED resembles the SEDs for mock galaxies with falling SFH in Figure 2 of analysis by \cite{Carnall2019}. The SED shape is used to constrain parameters such as metallicity, sSFR, and dust attenuation \citep{Conroy2013}. The differences in stellar mass and metallicity from our analysis suggest that the SED from the parametric model offsets from the one in the \texttt{continuity\_sfh model}. However, they both agree with the exponentially declining SFH shown in Figure \ref{fig:sfh}. 

In addition, the mass-weighted age is also dependent on the shape of the SED. We compare the average stellar population age with published results that range widely from 4 -- 11 Gyr \citep{Amblard2017}. Our results most closely resemble the age of $8.0 \pm 3.5$ Gyr calculated by \cite{Denicol2005}. Age is often known to be degenerate with dust unless spectroscopy or FIR data is available that can robustly measure the dust attenuation \citep{Conroy2013}. We obtained a 3-$\sigma$ upper limit on $A_V$ in both models. The results illustrate that NGC 2768 is indeed, an old and non-dusty environment with a strong constraint of $A_V$ $<$ 0.004 mag derived from the nonparametric model. 

As mentioned in Section \ref{sec:SED Modeling}, we included dust emission in both models for NGC 2768 in order to better fit IR photometry. The observed broadband photometry from both SEDs are consistent with the model photometry. We also observe absorption lines from the model spectra which is typical for a quiescent galaxy with sparse star formation.

\subsection{Supernova Rates} \label{sec:snrates}
We can quantitatively predict SN rates in our galaxies from the aforementioned SFHs using known forms of the supernova delay-time distribution (DTD), defined as the SN rate per unit mass of formed stars as a function of time elapsed since a hypothetical `burst' of star-formation \citep{MaozMan2012}. Theoretically, the DTD depends sensitively on the stellar evolution channels of the progenitor stars \citep{Mennekens2010, Toonen2012, Eldridge2017, Zapartas2017}, and has been extensively used to validate the progenitor channels of Type Ia SNe \citep{Totani2008, Maoz2010, Maoz2010a, Maoz2011, Maoz2012b, Maoz2014, Graur2014, Maoz2017}. While measuring a Ca-rich transient DTD is not feasible with our small, heterogenous sample, we can use existing constraints on the Type Ia and core-collapse SN DTD, convolve them with our measured SFHs, and roughly estimate the probability that the Ca-rich transients are from a Type Ia-like or core-collapse-like progenitor origin.

We restrict this analysis to the nine galaxies with nonparametric SFHs in Table \ref{tab:np results}, since the parametric SFHs may significantly underestimate the recent star-formation history and CC SN rates for star-forming galaxies. For each galaxy's SFH ($\dot{M}(t)$ in units of M$_{\odot}$/yr), the SN rate in a galaxy observed at a cosmic time (t) can be evaluated using:

\begin{equation*}
    R_{sn}(t) = \int_{0}^{t} \dot{M}(t - t_d) \Psi(t_d) \mathrm{d} t_d
\end{equation*}

\noindent{where $\Psi(t_d) $ is the delay-time distribution (DTD) for either CC or Type Ia SN as a function of delay-time $t_d$ \citep{Maoz2010}, and $R_{sn}(t)$ is the SN rate at a given cosmic time $t$. For each age bin ``j'' and corresponding SFR from the nonparametric model, the integral can be spread as:}

\begin{equation*}
    R_{sn} = \sum_{j=1}^{N_{age}} \dot{M}_j \int_{t_{jl}}^{t_{ju}} \Psi(t_d) \mathrm{d} t_d
\end{equation*}

\noindent{where $N_{age}$ is the number of age bins in the SFH, and $t_{ju}, t_{jl}$ are the upper and lower bounds of each age bin. Since $\dot{M}$ is constant in each age bin, it appears outside the integral.}

For core-collapse SNe, we use the form of $\Psi(t_d)$ in Eqn A2 of \cite{Zapartas2017}. The DTD is similar to the case of single stellar evolution for $t_d<40$ Myrs, with the exception of a `delayed' core-collapse SN channel at $t_d=40-200$ Myrs produced by stellar mergers of 4-8 M$_{\odot}$ stars (this channel however only accounts for 7-22$\%$ of CC SN population). For Type Ia SNe, we assume the form of $\Psi(t_d)$ constrained by \cite{MaozMan2012} from SN Ia surveys, i.e. $\Psi(t_d) = 4 \times 10^{-13}\ \mathrm{SN\ yr^{-1}\  M_{\odot}^{-1}} (t_d/1\ \mathrm{Gyr})^{-1}$ for $t_d>40$ Myrs. With the measured SN rates of core-collapse ($R_{cc}$) and Type Ia ($R_{Ia}$), we quantify the probability that the Ca-rich transient in each galaxy is a SN Ia or core-collapse event as $p_{cc} = R_{cc}/(R_{cc}+R_{Ia})$ and $p_{Ia} = 1 - p_{cc}$. These quantities will have uncertainties due to the uncertainties in the measured star-formation histories. We propagate these uncertainties using a Monte Carlo method. For each galaxy, we generate $5\times 10^4$ randomized SFHs assuming a log-normal distribution of SFRs for each age-bin. We then calculate the SN rates and probabilities for each of these randomized SFHs, and measure the median, 16th and 84th percentiles of the distributions.

The CC and SNe Ia rates and their probabilities for each of the nine host galaxies are summarized in Table \ref{tab:SN Rates}. The SNe Ia rate from quiescent galaxies (NGC 2768, NGC 4589, and NGC 2272) are nearly 20 -- 30 times higher than CC SN rates, with less than $5$\% probability respectively that the Ca-rich transients originate from massive progenitor systems. The star-forming galaxies in our nonparametric SFH sample, as expected, show a higher probability of producing CC SNe, excluding the host of PTF10hcw which has a relatively lower SFR.

Because only nine Ca-rich transient host galaxies could be modeled with nonparametric SFH, our ability to draw conclusions about the entire population of Ca-rich transients is limited.  Nonetheless, our analysis uncovered three examples where a core-collapse origin is improbable ($< 5\%$). Thus, our work supports the notion that the progenitor population of Ca-rich transients do not come exclusively from core-collapse explosions, and must either be only from WD stars or a mixed population of WD stars with other channels, potentially including massive star explosions. The fraction of CC vs.\ Type Ia origin can be quantified more accurately in subsequent studies with larger, homogeneous Ca-rich transient sample and with spatially-resolved star-formation histories of the galaxies, giving a more accurate age distribution of the SN progenitor \citep{Maoz2010, Chen2021, Sarbadhicary2021}. Improvements can also be made by utilizing better DTDs of potential progenitor channels. Instead of using the DTDs of SNe Ia and CC as proxies of Ca-rich transients as we have here, it would be advantageous to employ DTDs that are derived directly from theoretical models of proposed Ca-rich progenitor channels.

\section{Conclusions} 
In this paper we have presented multiwavelength observations of the host galaxies for all known Ca-rich transients and an uniform treatment of the stellar population modeling using \textsc{prospector}. The use of two separate models (parametric vs.\ nonparametric) allows us to compare and evaluate the effect of assumptions made in the stellar population properties. We reach the following conclusions:
\begin{itemize}

  \item From our SED fitting of all host galaxies using the parametric model, the average mass-weighted age for E-type host galaxies is 5.95 Gyr and 3.85 Gyr for S-type galaxies.

  \item $\approx$42$\%$ of all hosts, including two S-type host galaxies, can be considered as \textit{red and dead} (e.g. sSFR $<10^{-11}$\,yr$^{-1}$; \citealt{Fontana}), deeming them quiescent. This suggests that star formation has been suppressed in these host environments and that they are not associated with young massive stars.
  
  \item We performed nonparametric SED fitting using the \texttt{continuity\_sfh model} for the host galaxies of SN2000ds, SN2001co, SN2003dr, SN2005cz, SN2012hn, SN2016hnk, SN2019ehk, PTF10hcw and iPTF15eqv (three E-type and six S-type). We find that the mass-weighted ages and the stellar populations of the host galaxies are consistently older and more massive, respectively, compared to results from the parametric model. Further, the majority of stellar population mass was formed in the first few Gyr of the hosts' lifetimes and therefore the $t_m$ weighs them more heavily than stars that are born in later years. 
  
  \item We construct the SFH of our nonparametric fits and find that the E-type galaxies exhibit exponentially declining SFH as expected for older stellar populations, with no evidence of recent ($<100$ Myr) bursts of star formation. The S-type galaxies show recent star formation in the last 30 Myr (except NGC 2639) with an overall significantly higher SFR compared to the E-type hosts.

  \item Among the nine galaxies we are able to perform a detailed (nonparametric) star formation history, we estimate a $<5$\% probability for three separate host galaxies that the Ca-rich transients originated from a core collapse explosion, taking into account known forms of the Type Ia and core-collapse SN delay-time distributions. Thus, we find it very unlikely that the progenitor population of Ca-rich transients originates exclusively from core-collapse explosions.
  
 \end{itemize}

Our work  supports  the  notion  that  the  progenitors  of  Ca-rich transients  must  either  be  only from white dwarf stars or a mixed population of white dwarf stars with other channels, potentially including massive star explosions. Moving forward, the addition of spectra and/or narrow-band photometry to our fits can improve the mass recovery of stellar populations and better constrain age, metallicity, and early star formation activities \citep{Conroy2013, Lower2020}. The availability of more broadband photometry will also be advantageous for nonparametric techniques that offer SFH reconstruction methods with unbiased priors. Especially desirable is explosion site spectroscopy of larger samples of Ca-rich host galaxies, since parameters inferred for the hosts may not accurately reflect those of the Ca-rich transient environments, given that some are far offset from the optical radius. DTD models of proposed Ca-rich transient progenitors are also needed to quantitatively predict rates more accurately. 

Our method of photometry-only population inferencing will remain a primary way of investigating numerous Ca-rich transient host galaxies at high redshift. This is important because there is the concern that the present distribution of early-type and late-type host galaxies of Ca-rich is influenced by selection effects \citep{Lunnan}, and larger numbers of Ca-rich transients discovered in un-targeted surveys are needed. The Vera Rubin Observatory, which will discover thousands Ca-rich transients over the course of its ten year all-sky transient survey, has the potential to mitigate these selection biases and provide quality photometric data from which to perform nonparametric fits and best constrain the presently debated ratio of dwarf vs.\ massive star progenitor populations. 

\section{Acknowledgements}
D.~M.\ acknowledges NSF support from grants PHY-1914448 and AST-2037297. We are grateful to Samir Salim for useful discussions of photometric data and catalog usage. Y.~D.\ thanks Niharika Sravan, Mark Linvill, Alexander M. Warner, and Nabeel Rehemtulla for programming and plotting assistance. W.~J-G is supported by the National Science Foundation Graduate Research Fellowship Program under Grant No.~DGE-1842165. W.~J-G\ acknowledges support through NASA grants in support of {\it Hubble Space Telescope} programs GO-16075 and GO-16500. S.K.S.\ acknowledges support from NSF AST-1907790 and the Packard Foundation. This research has made use of data products from the Two Micron All Sky Survey, which is a joint project of the University of Massachusetts and the Infrared Processing and Analysis Center/California Institute of Technology, funded by the National Aeronautics and Space Administration and the National Science Foundation.
We also used the NASA/IPAC Extragalactic Database, which is funded by the National Aeronautics and Space Administration and operated by the California Institute of Technology.

This research is based on observations made with the NASA/ESA Hubble Space Telescope obtained from the Space Telescope Science Institute, which is operated by the Association of Universities for Research in Astronomy, Inc., under NASA contract NAS 5–26555.

\vspace{5mm}

\software{\texttt{Prospector} \citep{Leja2017}, \texttt{Python-fsps} \citep{FSPS, FSPSpython}, \texttt{dynesty} \citep{2020MNRAS.493.3132S}, \texttt{SAOImage DS9} \citep{ds9}}

\appendix

\setlength\LTright{3cm}
\begin{longtable}{ccccccc}
    \caption{Ca-rich Transients Host Galaxy Photometry
    \label{tab:photometry}} \\
	\hline
	\hspace{-2em} Host\footnote[1]{2MASX J22 = 2MASX J22465295+2138221, WISEA = WISEA J130109.43+280159.1, and 2MASX J00 = 2MASX J00505254+2722432} & Transient & \multicolumn{1}{p{1.5cm}}{\centering RA \\ (J2000)} & \multicolumn{1}{p{1.5cm}}{\centering Dec \\ (J2000)} &  Filter & $m_{AB}$\footnote{Corrected for Galactic Extinction (Schlegel et al. 1998) except W3 and W4}$^{,}$\footnote{An uncertainty of 0.005 was imposed on SDSS and DESI photometry.} & Refs.\footnote[4]{(1) \cite{2019ApJS..244...24L}, (2) \cite{Bai2015}, (3) SDSS, (4) 2MASS, (5) DESI, (6) GALEX, (7) Pan-STARRS DR2} \\ \hline    
	
    \hspace{-2em}NGC 2768 & SN 2000ds & $09^h11^m37.50^s$ & +60$^{\circ}$02'14.0'' & \multicolumn{1}{p{2cm}}{\centering FUV \\ NUV \\ B \\ I \\ J \\ H \\ K \\ W1 \\ W2 \\ W3 \\ W4} & \multicolumn{1}{p{3cm}}{\centering $16.578 \pm 0.003$ \\ $15.140 \pm 0.001$ \\ $10.460 \pm 0.020$ \\ $9.550 \pm 0.030$ \\ $8.800 \pm 0.020$ \\ $8.600 \pm 0.030$ \\ $8.830 \pm 0.030$ \\ $9.647 \pm 0.0001$ \\ $10.124 \pm 0.0003$ \\ $11.1950 \pm 0.009$ \\ $12.556 \pm 0.111$} & \multicolumn{1}{p{3cm}}{\centering 1 \\ 1 \\ 2 \\ 2 \\ 2 \\ 2 \\ 2 \\ 1 \\ 1 \\ 1 \\ 1} \\
		
	\hspace{-2em}NGC 5559 & SN 2001co & $14^h19^m12.79^s$ & +24$^{\circ}$47'55.42'' & \multicolumn{1}{p{2cm}}{\centering FUV \\ NUV \\ g \\ r \\ i \\ z \\ J \\ H \\ K \\ W1 \\ W2 \\ W3 \\ W4} & \multicolumn{1}{p{3cm}}{\centering $17.551 \pm 0.001$ \\ $16.960 \pm 0.001$ \\ $14.326 \pm 0.005$ \\ $13.516 \pm 0.005$ \\ $13.059 \pm 0.005$ \\ $12.781 \pm 0.005$ \\ $12.348 \pm 0.020$ \\ $12.209 \pm 0.020$ \\ $12.209 \pm 0.031$ \\ $12.758 \pm 0.0003$ \\ $13.259 \pm 0.001$ \\ $11.741 \pm 0.003$ \\ $11.434 \pm 0.008$ }   & \multicolumn{1}{p{3cm}}{\centering 1 \\ 1 \\ 3 \\ 3 \\ 3 \\ 3 \\ 4 \\ 4 \\ 4 \\ 1 \\ 1 \\ 1 \\ 1} \\

	\hspace{-2em}NGC 2207 & SN 2003H & $06^h16^m22.03^s$ & -21$^{\circ}$22'21.60'' & \multicolumn{1}{p{2cm}}{\centering FUV \\ NUV \\ B \\ I \\ W1 \\ W2 \\ W3 \\ W4} & \multicolumn{1}{p{3cm}}{\centering $12.803 \pm 0.001$ \\ $ 12.439 \pm 0.0001$ \\ $11.700 \pm 0.300$ \\ $10.090 \pm 0.100$\\ $10.045 \pm 0.0001$ \\ $10.545 \pm 0.0003$ \\ $8.629 \pm 0.001$ \\ $7.964 \pm 0.001$} & \multicolumn{1}{p{3cm}}{\centering 1 \\ 1 \\ 2 \\ 2 \\ 1 \\ 1 \\ 1 \\ 1} \\
		
	\hspace{-2em}UGC 6934 & SN 2003dg & $11^h57^m31.83^s$ & -01$^{\circ}$15'10,58'' & \multicolumn{1}{p{2cm}}{\centering FUV \\ NUV \\ J \\ H \\ K \\} & \multicolumn{1}{p{3cm}}{\centering $18.00 \pm 0.01$ \\ $17.59 \pm 0.01$ \\ $13.49 \pm 0.04$ \\ $13.23 \pm 0.05$ \\ $13.44 \pm 0.08$}  & \multicolumn{1}{p{3cm}}{\centering 2 \\ 2 \\ 2 \\ 2 \\ 2}  \\
	
	\hspace{-2em}NGC 5714 & SN 2003dr & $14^h38^m11.52^s$ & +46$^{\circ}$38'17.70'' & \multicolumn{1}{p{2cm}}{\centering FUV \\ NUV \\ g \\ r \\ i \\ z \\ J \\ H \\ K \\ W1 \\ W2 \\ W3 \\ W4} & \multicolumn{1}{p{3cm}}{\centering $16.218 \pm 0.003$ \\ $15.952 \pm 0.002$ \\ $13.924 \pm 0.005$ \\ $13.149 \pm 0.005$ \\ $12.749 \pm 0.005$ \\ $12.448 \pm 0.005$ \\ $12.032 \pm 0.022$ \\ $11.722 \pm 0.031$ \\ $11.902 \pm 0.035$ \\ $12.294 \pm 0.0004$ \\ $12.810 \pm 0.001$ \\ $11.364 \pm 0.004$ \\ $11.082 \pm 0.012$}  & \multicolumn{1}{p{3cm}}{\centering 1 \\ 1 \\ 3 \\ 3 \\ 3 \\ 3 \\ 3 \\ 3\\ 3 \\ 1 \\ 1 \\ 1 \\ 1} \\
		
	\hspace{-2em}NGC 1032 & SN 2005E & $02^h39^m23.64^s$ & +01$^{\circ}$05'37.64'' & \multicolumn{1}{p{2cm}}{\centering NUV \\ g \\ r \\ z \\ W1 \\ W2 \\ W3 \\ W4 } & \multicolumn{1}{p{3cm}}{\centering $16.229 \pm 0.018$ \\ $11.786 \pm 0.005$ \\ $10.955 \pm 0.005$ \\ $10.234 \pm 0.005$ \\ $10.869 \pm 0.0002$ \\ $11.495 \pm 0.001$ \\ $11.686 \pm 0.008$ \\ $11.661 \pm 0.026$}  & \multicolumn{1}{p{3cm}}{\centering 1 \\ 5 \\ 5 \\ 5 \\ 1 \\ 1 \\ 1 \\ 1 }\\
		
	\hspace{-2em}NGC 4589 & SN 2005cz & $12^h37^m24.99^s$ & +74$^{\circ}$11'30.92'' & \multicolumn{1}{p{2cm}}{\centering FUV \\ NUV \\ J \\ H \\ K \\ W1 \\ W2 \\ W3 \\ W4} & \multicolumn{1}{p{3cm}}{\centering $17.614 \pm 0.003$ \\ $15.904 \pm 0.001$ \\ $9.742 \pm 0.016$ \\ $9.516 \pm 0.016$ \\ $9.755 \pm 0.017$ \\ $10.312 \pm 0.0001$ \\ $10.997 \pm 0.0002$ \\ $12.207 \pm 0.009$ \\ $12.453 \pm 0.039$}  & \multicolumn{1}{p{3cm}}{\centering 1 \\ 1 \\ 4 \\ 4 \\ 4 \\ 1 \\ 1 \\ 1 \\ 1} \\
		
    \hspace{-2em}NGC 1129 & SN 2007ke & $02^h54^m27.38^s$ & +41$^{\circ}$34'46.52'' & \multicolumn{1}{p{2cm}}{\centering NUV \\ u \\ g \\ r \\ i \\ z} & \multicolumn{1}{p{3cm}}{\centering $18.400 \pm 0.150$ \\ $14.526 \pm 0.005$ \\ $12.471 \pm 0.005$ \\ $11.589 \pm 0.005$ \\ $11.144 \pm 0.005$ \\ $10.873 \pm 0.005$} & \multicolumn{1}{p{3cm}}{\centering 6 \\ 3 \\ 3 \\ 3 \\ 3 \\ 3}  \\
		
	\hspace{-2em}CGCG 170-011 & SN 2010et & $17^h16^m52.18^s$ & +31$^{\circ}$35'02.88'' & \multicolumn{1}{p{2cm}}{\centering g \\ r \\ i \\ z \\ J \\ H \\ K } & \multicolumn{1}{p{3cm}}{\centering $14.784 \pm 0.005$ \\ $14.040 \pm 0.005$ \\ $13.619 \pm 0.005$ \\ $13.355 \pm 0.005$ \\ $13.499 \pm 0.041$ \\ $13.290 \pm 0.026$ \\ $13.558 \pm 0.017$}  & \multicolumn{1}{p{3cm}}{\centering 3 \\ 3 \\ 3 \\ 3 \\ 4 \\ 4 \\ 4} \\
		
	\hspace{-2em}NGC 2272 & SN 2012hn & $06^h42^m41.30^s$ & -27$^{\circ}$27'34.20'' & \multicolumn{1}{p{2cm}}{\centering FUV \\ NUV \\ J \\ H \\ K \\ W1 \\ W2 \\ W3 \\ W4 } & \multicolumn{1}{p{3cm}}{\centering $17.667 \pm 0.021$ \\ $16.227 \pm 0.005$ \\ $10.581 \pm 0.023$ \\ $10.402 \pm 0.027$ \\ $10.648 \pm 0.034$ \\ $11.178 \pm 0.0002$ \\ $11.861 \pm 0.001$ \\ $12.609 \pm 0.014$ \\ $14.231 \pm 0.227$}  & \multicolumn{1}{p{3cm}}{\centering 1 \\ 1 \\ 4 \\ 4 \\ 4 \\ 1 \\ 1 \\ 1 \\ 1 } \\
		
	\hspace{-2em}MGC-01-06-070 & SN 2016hnk & $02^h13^m15.79^s$ & -07$^{\circ}$39'42.70'' & \multicolumn{1}{p{2cm}}{\centering FUV \\ NUV \\ g \\ r \\ i \\ z \\ W1 \\ W2 \\ W3 \\ W4} & \multicolumn{1}{p{3cm}}{\centering $16.182 \pm 0.001$ \\ $15.825 \pm 0.001$ \\ $14.339 \pm 0.005$ \\ $13.517 \pm 0.005$ \\ $13.095 \pm 0.005$ \\ $12.781 \pm 0.005$ \\ $12.723 \pm 0.001$ \\ $13.401 \pm 0.002$ \\ $12.926 \pm 0.014$ \\ $12.603 \pm 0.036$}  & \multicolumn{1}{p{3cm}}{\centering 1 \\ 1 \\ 3 \\ 3 \\ 3 \\ 3 \\ 1 \\ 1 \\ 1 \\ 1} \\
		
	\hspace{-2em}NGC 4321 & SN 2019ehk & $12^h22^m54.831^s$ & +15$^{\circ}$49'18.54'' & \multicolumn{1}{p{2cm}}{\centering FUV \\ NUV \\ J \\ H \\ K \\ W1 \\ W2 \\ W3 \\ W4} & \multicolumn{1}{p{3cm}}{\centering $12.547 \pm 0.0001$ \\ $11.998 \pm 0.00004$ \\ $8.593 \pm 0.016$ \\ $8.463 \pm 0.018$ \\ $8.650 \pm 0.018$ \\ $8.950 \pm 0.0001$ \\ $9.471 \pm 0.0002$ \\ $7.808 \pm 0.001$ \\ $7.412 \pm 0.002$} & \multicolumn{1}{p{3cm}}{\centering 1 \\ 1 \\ 4 \\ 4 \\ 4 \\ 1 \\ 1 \\ 1 \\ 1 } \\
		
	\hspace{-2em} 2MASX J22 & PTF 09dav & $22^h46^m52.95^s$ & +21$^{\circ}$38'21.90'' & \multicolumn{1}{p{2cm}}{\centering g \\ r \\ i \\ y \\ J \\ H \\ K} & \multicolumn{1}{p{3cm}}{\centering $17.931 \pm 0.064$ \\ $17.588 \pm 0.087$ \\ $17.227 \pm 0.064$ \\ $16.621 \pm 0.123$ \\ $15.329 \pm 0.083$ \\ $15.109 \pm 0.075$ \\ $15.374 \pm 0.149$} & \multicolumn{1}{p{3cm}}{\centering 7 \\ 7 \\ 7 \\ 7 \\ 4 \\ 4 \\ 4} \\	
		
	\hspace{-2em}NGC 2639 & PTF 10hcw & $08^h43^m38.08^s$ & +50$^{\circ}$12'20.00'' & \multicolumn{1}{p{2cm}}{\centering FUV \\ NUV \\ g \\ r \\ i \\ z \\ J \\ H  \\ K \\ W1 \\ W2 \\ W3 \\ W4} & \multicolumn{1}{p{3cm}}{\centering $16.961 \pm 0.001$ \\ $15.908 \pm 0.001$ \\ $12.100 \pm 0.005$ \\ $11.225 \pm 0.005$ \\ $10.871 \pm 0.005$ \\ $10.515 \pm 0.005$ \\ $10.392 \pm 0.016$ \\ $10.159 \pm 0.016$ \\ $10.340 \pm 0.016$ \\ $10.963 \pm 0.0001$ \\ $11.589 \pm 0.0004$ \\ $11.101 \pm 0.003$ \\ $10.796 \pm 0.008$}  & \multicolumn{1}{p{3cm}}{\centering 1 \\ 1 \\ 3 \\ 3 \\ 3 \\ 3 \\ 4 \\ 4 \\ 4 \\ 1 \\ 1 \\ 1 \\ 1} \\
		
    \hspace{-2em}IC 3956 & PTF 11bij & $12^h58^m56.38^s$ & +37$^{\circ}$23'53.44'' & \multicolumn{1}{p{2cm}}{\centering J \\ H \\ K \\ W1 \\ W2} & \multicolumn{1}{p{3cm}}{\centering $13.926 \pm 0.030$ \\ $13.707 \pm 0.037$ \\ $13.858 \pm 0.050$ \\ $14.337 \pm 0.005$ \\ $14.987 \pm 0.005$}  & \multicolumn{1}{p{3cm}}{\centering 4 \\ 4 \\ 4 \\ 5 \\ 5} \\
		
	\hspace{-2em}NGC 7265 & PTF 11kmb & $22^h22^m27.44^s$ & +36$^{\circ}$12'34.58'' & \multicolumn{1}{p{2cm}}{\centering NUV \\ g \\ r \\ i} & \multicolumn{1}{p{3cm}}{\centering $18.500 \pm 0.066$ \\ $14.471 \pm 0.016$ \\ $14.351 \pm 0.038$ \\ $13.527 \pm 0.050$}  & \multicolumn{1}{p{3cm}}{\centering 6 \\ 7 \\ 7 \\ 7} \\
		
	\hspace{-2em} WISEA & PTF 12bho & $13^h01^m09.44^s$ & +28$^{\circ}$01'59.21'' & \multicolumn{1}{p{2cm}}{\centering g \\ r \\ i \\ z \\ y} & \multicolumn{1}{p{3cm}}{\centering $17.409 \pm 0.007$ \\ $16.711 \pm 0.027$ \\ $16.399 \pm 0.059$ \\ $16.189 \pm 0.027$ \\ $15.937 \pm 0.022$} & \multicolumn{1}{p{3cm}}{\centering 7 \\ 7 \\ 7 \\ 7 \\ 7} \\
		
	\hspace{-2em}NGC 3430 & iPTF 15eqv & $10^h52^m11.40^s$ & +32$^{\circ}$57'01.56'' & \multicolumn{1}{p{2cm}}{\centering FUV \\ NUV \\ g \\ r \\ i \\ z \\ J \\ H \\ K \\ W1 \\ W2 \\ W3 \\ W4} & \multicolumn{1}{p{3cm}}{\centering $14.089 \pm 0.0002$ \\ $13.769 \pm 0.0002$ \\ $12.570 \pm 0.005$ \\ $11.795 \pm 0.005$ \\ $11.391 \pm 0.005$ \\ $11.185 \pm 0.005$ \\ $10.836 \pm 0.014$ \\ $10.651 \pm 0.018$ \\ $10.869 \pm 0.024$ \\ $11.285 \pm 0.0003$ \\ $11.867 \pm 0.001$ \\ $10.119 \pm 0.003$ \\ $9.817 \pm 0.007$}  & \multicolumn{1}{p{3cm}}{\centering 1 \\ 1 \\ 3 \\ 3 \\ 3 \\ 3 \\ 4 \\ 4 \\ 4 \\ 1 \\ 1 \\ 1 \\ 1} \\
	
	\hspace{-2em} 2MASX J00 & iPTF 16hgs & $00^h50^m52.55^s$ & +27$^{\circ}$22'42.78'' & \multicolumn{1}{p{2cm}}{\centering g \\ r \\ J \\ H \\ K} & \multicolumn{1}{p{3cm}}{\centering $17.468 \pm 0.041$ \\ $17.015 \pm 0.036$ \\ $16.227 \pm 0.059$ \\ $16.119 \pm 0.037$ \\ $16.148 \pm 0.024$} & \multicolumn{1}{p{3cm}}{\centering 7 \\ 7 \\ 4 \\ 4 \\ 4} \\

\end{longtable}
\label{table:1}

\bibliography{refs}

\end{CJK*}
\end{document}